\documentclass[journal, pdftex, draftclsnofoot, onecolumn, 12pt, twoside]{IEEEtran}
\usepackage{amsfonts}
\usepackage[ruled,vlined]{algorithm2e}
\usepackage{pgfplots}
\usepackage{pgfplotstable}

\pgfplotsset{compat=newest}
\usepackage[noend]{algpseudocode}
\usepackage{amssymb}
\usepackage{cite}
\usepackage[cmex10]{amsmath}
\ifCLASSINFOpdf
\graphicspath{{./Images/}}
\usepackage{array}
\usepackage[caption=false, font=footnotesize]{subfig}

\DeclareMathOperator*{\argmax}{arg\,max}
\ifCLASSOPTIONcompsoc
    \usepackage[caption=false, font=normalsize, labelfont=sf, textfont=sf]{subfig}
\else
\usepackage[caption=false, font=footnotesize]{subfig}
\fi

\ifCLASSINFOpdf
\else
\fi

\IEEEoverridecommandlockouts
\usepackage{xcolor}
\usepackage{amssymb}
\def\BibTeX{{\rm B\kern-.05em{\sc i\kern-.025em b}\kern-.08em
    T\kern-.1667em\lower.7ex\hbox{E}\kern-.125emX}}

\newtheorem{theorem}{Theorem}

\newtheorem{Remark}{Remark}
\IEEEoverridecommandlockouts

\begin{document}
\title{Effective Communications: A Joint Learning and Communication Framework for Multi-Agent Reinforcement Learning over Noisy Channels}
%
\author{Tze-Yang Tung, Szymon Kobus, Joan Pujol Roig, Deniz G\"und\"uz \\
Information Processing and Communications Laboratory (IPC-Lab)\\
Dept. of Electrical and Electronic Engineering, Imperial College London, UK
\thanks{This work was supported in part by the European Research Council (ERC) Starting Grant BEACON (grant agreement no. 677854) and by the UK EPSRC (grant no. EP/T023600/1).}
\thanks{An earlier version of this work was presented at the IEEE Global Communications Conference (GLOBECOM) in December 2020 \cite{Roig:Globecom:20}.}
}

\maketitle
\vspace{-1.3cm}
\begin{abstract}
We propose a novel formulation of the ``effectiveness problem'' in communications, put forth by Shannon and Weaver in their seminal work \cite{ShannonWeaver49}, by considering multiple agents communicating over a noisy channel in order to achieve better coordination and cooperation in a multi-agent reinforcement learning (MARL) framework. 
Specifically, we consider a multi-agent partially observable Markov decision process (MA-POMDP), in which the agents, in addition to interacting with the environment can also communicate with each other over a noisy communication channel. 
The noisy communication channel is considered explicitly as part of the dynamics of the environment and the message each agent sends is part of the action that the agent can take. 
As a result, the agents learn not only to collaborate with each other but also to communicate ``effectively'' over a noisy channel. 
This framework generalizes both the traditional communication problem, where the main goal is to convey a message reliably over a noisy channel, and the ``learning to communicate'' framework that has received recent attention in the MARL literature, where the underlying communication channels are assumed to be error-free. We show via examples that the joint policy learned using the proposed framework is superior to that where the communication is considered separately from the underlying MA-POMDP. 
This is a very powerful framework, which has many real world applications, from autonomous vehicle planning to drone swarm control, and opens up the rich toolbox of deep reinforcement learning for the design of multi-user communication systems. 
\end{abstract}


\section{Introduction}
\label{sec:intro}  

Communication is essential for our society. 
Humans use language to communicate ideas, which has given rise to complex social structures, and scientists have observed either gestural or vocal communication in other animal groups, complexity of which increases with the complexity of the social structure of the group \cite{tomasello_origins_2010}. 
Communication helps to achieve complex goals by enabling cooperation and coordination \cite{ackley:alife4, KamChuen:AGENTS:01}. 
Advances in our ability to store and transmit information over time and long distances have greatly expanded our capabilities, and allows us to turn the world into the connected society that we observe today.
Communication technologies are at the core of this massively complex system. 

Communication technologies are built upon fundamental mathematical principles and engineering expertise. 
The fundamental quest in the design of these systems have been to deal with various imperfections in the communication channel (e.g., noise and fading) and the interference among transmitters. 
Decades of research and engineering efforts have produced highly advanced networking protocols, modulation techniques, waveform designs and coding techniques that can overcome these challenges quite effectively. 
However, this design approach ignores the aforementioned core objective of communication in enabling coordination and cooperation. 
To some extent, we have separated the design of a communication network that can reliably carry signals from one point to another from the `language' that is formed to achieve coordination and cooperation among agents. 

This engineering approach was also highlighted by Shannon and Weaver in \cite{ShannonWeaver49} by organizing the communication problem into three ``levels": They described level A as the \textit{technical problem}, which tries to answer the question ``How accurately can the symbols of communication be transmitted?". 
Level B is referred to as the \textit{semantic problem}, and asks the question ``How precisely do the transmitted symbols convey the desired meaning?". 
Finally, Level C, called the \textit{effectiveness problem}, strives to answer the question ``How effectively does the received meaning affect conduct in the desired way?". 
As we have described above, our communication technologies mainly deal with Level A, ignoring the semantics or the effectiveness problems. 
This simplifies the problem into the transmission of a discrete message or a continuous waveform over a communication channel in the most reliable manner. 
The semantics problem deals with the meaning of the messages, and is rather abstract. 
There is a growing interest in the semantics problem in the recent literature \cite{Guler:TCCN:18, popovski2019semanticeffectiveness, kountouris2020semanticsempowered, xie2020deep, strinati20206g}.
However, these works typically formulate the semantics as an end-to-end joint source-channel coding problem, where the reconstruction objective can be distortion with respect to the original signal \cite{bourtsoulatze_deep_2018, weng2020semantic}, or a more general function that can model some form of `meaning' \cite{Guler:TCCN:18, sreekumar_distributed_2020, Jankowski:JSAC:21, Gunduz:CL:20}, which goes beyond reconstructing the original signal\footnote{To be more precise, remote hypothesis testing, classification, or retrieval problems can also be formulated as end-to-end joint source-channel coding problems, albeit with a non-additive distortion measure.}. 

\begin{figure}
    \centering
    \includegraphics[width=0.3\linewidth]{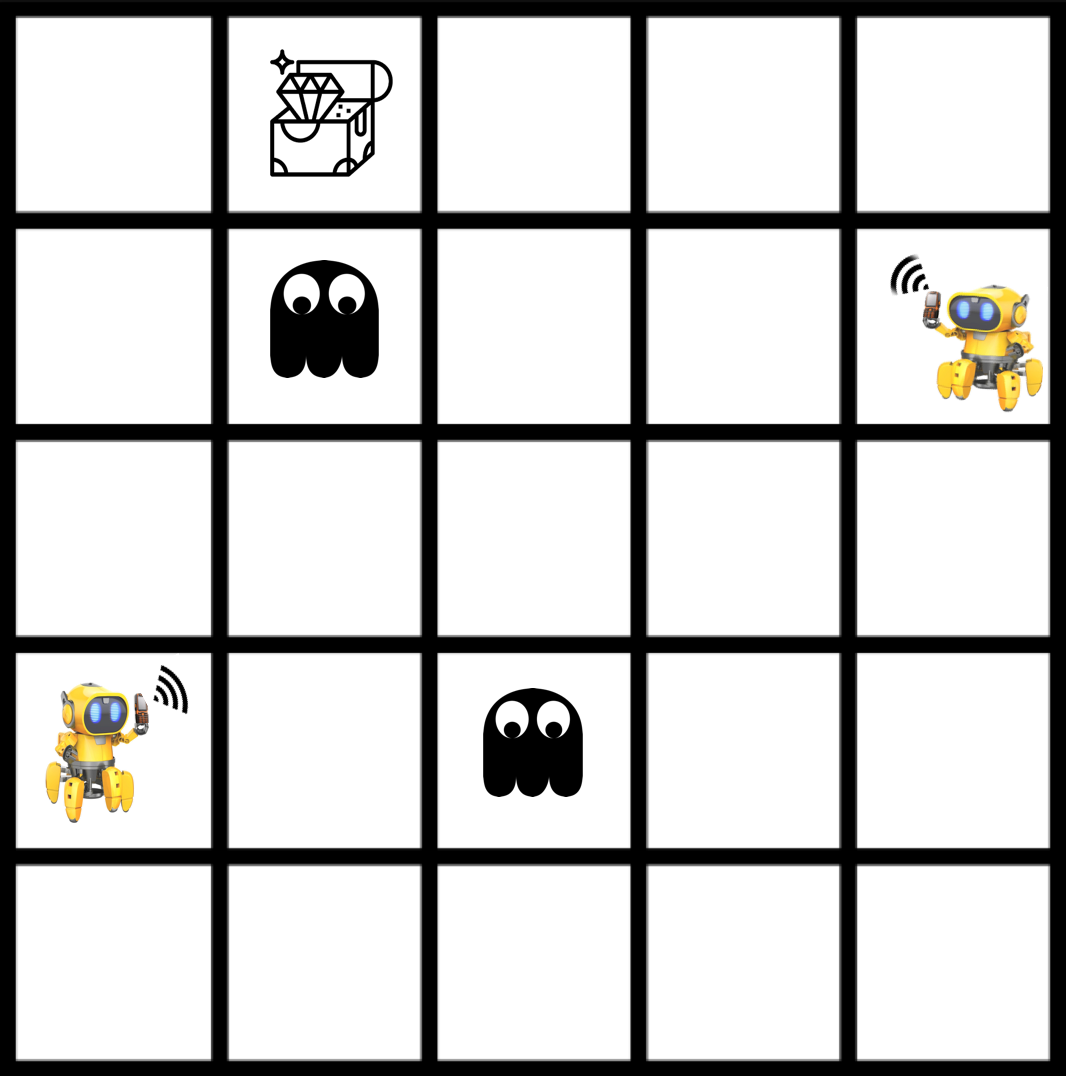}
    \caption{An illustration of a MARL problem with noisy communication between the agents, e.g., agents communicating over a shared wireless channel. The emerging communication scheme should not only allow the agents to better coordinate and cooperate to maximize their rewards, but also mitigate the adverse effects of the wireless channel, such as noise and interference.}
    \label{fig:MARLwComms}
\vspace{-0.8cm}
\end{figure}

In this paper, we deal with the `effectiveness problem', which generalizes the problems in both level A and level B. In particular, we formulate a multi-agent problem with noisy communications between the agents, where the goal of communications is to help agents better cooperate and achieve a common goal. See Fig. \ref{fig:MARLwComms} for an illustration of a multi-agent grid-world, where agents can communicate through noisy wireless links. 
It is well-known that multi-agent reinforcement learning (MARL) problems are notoriously difficult, and are a topic of continuous research. Originally, these problems were approached by treating each agent independently, as in a standard single-agent reinforcement learning (RL) problem, while treating other agents as part of the state of the environment. Consensus and cooperation are achieved through common or correlated reward signals. However, this approach leads to overfitting of policies due to limited local observations of each agent and it relies on other agents not varying their policies \cite{lanctot_unified_2017}. It has been observed that these limitations can be overcome by leveraging communication between the agents \cite{KamChuen:AGENTS:01, Balch:AR:94}. 

Recently, there has been significant interest in the \textit{emergence of communication} among agents within the RL literature \cite{foerster_learning_2016, jiang_learning_2018, jaques_social_2019, das_tarmac_2020}.
These works consider MARL problems, in which agents have access to a dedicated communication channel, and the objective is to learn a communication protocol, which can be considered as a `language' to achieve the underlying goal, which is typically translated into maximizing a specific reward function. 
This corresponds to Level C, as described by Shannon and Weaver in \cite{ShannonWeaver49}, where the agents change their behavior based on the messages received over the channel in order to maximize their reward. 
However, the focus of the aforementioned works is the emergence of communication protocols within the limited communication resources that can provide the desired impact on the behavior of the agents, and, unlike Shannon and Weaver, these works ignore the physical layer characteristics of the channel. 

Our goal in this work is to consider the effectiveness problem by taking into account both the channel noise and the end-to-end learning objective. 
In this problem, the goal of communication is not ``reproducing at one point either exactly or approximately a message selected at another point'' as stated by Shannon in \cite{ShannonWeaver49}, which is the foundation of the communication and information theoretic formulations that have been studied over the last seven decades. 
Instead, the goal is to enable cooperation in order to improve the objective of the underlying multi-agent game. As we will show later in this paper, the codes that emerge from the proposed framework can be very different from those that would be used for reliable communication of messages. 

We formulate this novel communication problem as a MARL problem, in which the agents have access to a noisy communication channel. More specifically, we formulate this as a multi-agent partially observable Markov decision process (POMDP), and construct RL algorithms that can learn policies that govern both the actions of the agents in the environment and the signals they transmit over the channel. A communication protocol in this scenario should aim to enable cooperation and coordination among agents in the presence of channel noise. Therefore, the emerging modulation and coding schemes must not only be capable of error correction/ compensation, but also enable agents to share their knowledge of the environment and/or their intentions. We believe that this novel formulation opens up many new directions for the design of communication protocols and codes that will be applicable in many multi-agent scenarios from teams of robots to platoons of autonomous cars  \cite{wang_networking_2019}, to drone swarm planning \cite{campion_uav_2018}.



We summarize the main contributions of this work as follows: 

\begin{enumerate}
    \item We propose a novel formulation of the ``effectiveness problem'' in communications, where agents communicate over a noisy communication channel in order to achieve better coordination and cooperation in a MARL framework. This can be interpreted as a \textit{joint communication and learning approach} in the RL context \cite{Gunduz:CL:20}. The current paper is an initial study of this general framework, focusing on scenarios that involve only point-to-point communications for simplicity. More involved multi-user communication and coordination problems will be the subject of future studies. 
    
    \item The proposed formulation generalizes the recently studied ``learning to communicate'' framework in the MARL literature \cite{foerster_learning_2016, jiang_learning_2018, jaques_social_2019, das_tarmac_2020}, where the underlying communication channels are assumed to be error-free. 
    This framework has been used to argue about the emergence of natural languages \cite{lazaridou_multi-agent_2017, lazaridou2020multiagent}; however, in practice, there is inherent noise in any communication medium, particularly in human/animal communications. Indeed, languages have evolved to deal with such noise. For example, Shannon estimated that the English language has approximately 75\% redundancy. 
    Such redundancy provides error correction capabilities. 
    Hence, we argue that the proposed framework better models realistic communication problems, and the emerging codes and communication schemes can help better understand the underlying structure of natural languages. 
    
    \item The proposed framework also generalizes communication problems at level A, which have been the target of most communication protocols and codes that have been developed in the literature. 
    Channel coding, source coding, as well as joint source-channel coding problems, and their multi-user extensions can be obtained as special cases of the proposed framework. 
    The proposed deep reinforcement learning (DRL) framework provides alternative approaches to the design of codes and communication schemes for these problems that can outperform existing ones. 
    We highlight that there are very limited practical code designs in the literature for most multi-user communication problems, and the proposed framework and the exploitation of deep representations and gradient-based optimization in DRL can provide a scalable and systematic methodology to make progress in these challenging problems. 
      
    \item We study a particular case of the proposed general framework as an example, which reduces to a point-to-point communication problem. 
    In particular, we show that any single-agent Markov decision process (MDP) can be converted into a multi-agent partially observable MDP (MA-POMDP) with a noisy communication link between the two agents.  
    We consider both the binary symmetric channel (BSC), the additive white Gaussian noise (AWGN) channel, and the bursty noise (BN) channel for the noisy communication link and solve the MA-POMDP problem by treating the other agent as part of the environment, from the perspective of one agent.
    We employ deep Q-learning (DQN) \cite{mnih_human-level_2015} and deep deterministic policy gradient (DDPG) \cite{lillicrap_continuous_2019} to train the agents.
    Substantial performance improvement is observed in the resultant policy over those learned by considering the cooperation and communication problems separately.

    \item We then present the joint modulation and channel coding problem as an important special case of the proposed framework. 
    In recent years, there has been a growing interest in using machine learning techniques to design practical channel coding and modulation schemes \cite{Nachmani:STSP:18, Dorner:Asilomar:17, Felix:SPAWC:18, bourtsoulatze_deep_2018, Kurka:JSAIT:20, aoudia_model-free_2019}. 
    However, with the exception of \cite{aoudia_model-free_2019}, most of these approaches assume that the channel model is known and differentiable, allowing the use of supervised training by directly backpropagating through the channel using the channel model. In this paper, we learn to communicate over an unknown channel solely based on the reward function by formulating it as a RL problem. The proposed DRL framework goes beyond the method employed in \cite{aoudia_model-free_2019}, which treats the channel as a random variable, and numerically approximates the gradient of the loss function. It is shown through numerical examples that the proposed DRL techniques employing DDPG \cite{lillicrap_continuous_2019}, and actor-critic \cite{konda_actor-critic_nodate} algorithms  significantly improve the block error probability (BLER) of the resultant code.
\end{enumerate}

\section{Related Works}
\label{sec:related_works}

The study of communication for multi-agent systems is not new \cite{wagner_progress_2016}. 
However, due to the success of deep neural networks (DNNs) for reinforcement learning (RL), this problem has received renewed interest in the context of DNNs \cite{lazaridou_multi-agent_2017} and deep RL (DRL) \cite{foerster_learning_2016, sukhbaatar_learning_2016, havrylov_emergence_2017}, where partially observable multi-agent problems are considered. 
In each case, the agents, in addition to taking actions that impact the environment, can also communicate with each other via a limited-capacity communication channel. 
Particularly, in \cite{foerster_learning_2016}, two approaches are considered: reinforced inter-agent learning (RIAL), where two centralized Q-learning networks learn to act and communicate, respectively, and differentiable inter-agent learning (DIAL), where communication feedback is provided via backpropagation of gradients through the channel, while the communication between agents is restricted during execution. 
Similarly, in \cite{wang_r-maddpg_2020,lowe2017maddpg}, the authors propose a \textit{centralized learning, decentralized execution} approach, where a central critic is used to learn the state-action values of all the agents and use those values to train individual policies of each agent. 
Although they also consider the transmitted messages as part of the agents' actions, the communication channel is assumed to be noiseless.

CommNet \cite{sukhbaatar_learning_2016} attempts to leverage communications in cooperative MARL by using multiple continuous-valued transmissions at each time step to make decisions for all agents. 
Each agent broadcasts its message to every other agent, and the averaged message received by each agent forms part of the input.
However, this solution lacks scalability as it depends on a centralized network by treating the problem as a single RL problem. 
Similarly, BiCNet \cite{peng_multiagent_2017} utilizes recurrent neural networks to connect individual agent's policy with a centralized controller aggregating the hidden states of each agent, acting as communication messages.

The reliance of the aforementioned works on a broadcast channel to communicate with all the agents simultaneously may be infeasible or highly inefficient in practice.
To overcome this limitation, in \cite{jiang_learning_2018}, the authors propose an attentional communication model that learns when communication is needed and how to integrate shared information for cooperative decision making.
In \cite{das_tarmac_2020}, directional communication between agents is achieved with a signature-based soft attention mechanism, where each message is associated to the target recipient. 
They also propose multi-stage communication, where multiple rounds of communication take place before an action is taken.

It is important to note that, with the exception of \cite{mostaani_learning-based_2019}, all of the prior works discussed above rely on error-free communication channels.
MARL with noisy communications is considered in  \cite{mostaani_learning-based_2019}, where two agents placed on a grid world aim to coordinate to step on the goal square simultaneously. 
However, for the particular problem presented in \cite{mostaani_learning-based_2019}, it can be shown that even if the agents are trained independently without any communication at all, the total discounted reward would still be higher than the average reward achieved by the scheme proposed in \cite{mostaani_learning-based_2019}.


\section{Problem Formulation}
\label{sec:problem_formulation}

We consider a multi-agent partially observable Markov decision process (MA-POMDP) with noisy communications. 
Consider first a Markov game with $N$ agents $(\mathcal{S}, \{\mathcal{O}_i\}_{i=1}^N, \{\mathcal{A}_i\}_{i=1}^N,$ $ P, r)$, where $\mathcal{S}$ represents all possible configurations of the environment and agents, 
$\mathcal{O}_i$ and $\mathcal{A}_i$ are the observation and action sets of agent $i$, respectively, $P$ is the transition kernel that governs the environment, and $r$ is the reward function. 
At each step $t$ of this Markov game, agent $i$ has a partial observation of the state $o_i^{(t)}\in \mathcal{O}_i$, and takes action $a_i^{(t)} \in \mathcal{A}_i$, $\forall i$. 
Then, the state of the MA-POMDP transitions from $s^{(t)}$ to $s^{(t+1)}$ according to the joint actions of the agents following the transition probability $P(s^{(t+1)}|s^{(t)}, \mathbf{a}^{(t)})$, where $\mathbf{a}^{(t)} = (a_1^{(t)}, \ldots, a_N^{(t)})$. 
Observations in the next time instant follow the conditional distribution $\mathrm{Pr}(o^{(t+1)}|s^{(t)}, \mathbf{a}^{(t)})$. 
While, in general, each agent can have a separate reward function, we consider herein the fully cooperative setting, where the agents receive the same team reward $r^{(t)} = r(s^{(t)}, \mathbf{a}^{(t)})$ at time $t$. 

In order to coordinate and maximize the total reward, the agents are endowed with a noisy communication channel, which is orthogonal to the environment.
That is, the environment transitions depend only on the environment actions, and the only impact of the communication channel is that the actions of the agents can now depend on the past received messages as well as the past observations and rewards. 
We assume that the communication channel is governed by the conditional probability distribution $P_c$, and we allow the agents to use the channel $M$ times at each time $t$. 
Here, $M$ can be considered as the \textit{channel bandwidth}. 
Let the signals transmitted and received by agent $i$ at time step $t$ be denoted by $\mathbf{m}_i^{(t)} \in \mathcal{C}_t^M$ and $\hat{\mathbf{m}}_i^{(t)} \in\mathcal{C}_r^M$, respectively, where $\mathcal{C}_t$ and $\mathcal{C}_r$ denote the input and output alphabets of the channel, which can be discrete or continuous.
We assume for simplicity that the input and output alphabets of the channel are the same for all the agents. Channel inputs and outputs at time $t$ are related through the conditional distribution $P_c\big(\hat{\mathbf{M}}^{(t)} | \mathbf{M}^{(t)} \big) =\mathrm{Pr}\big(\hat{\mathbf{M}} = \{\hat{\mathbf{m}}_i^{(t)}\}_{i=1}^N \big|\mathbf{M}=\{\mathbf{m}_i^{(t)}\}_{i=1}^N \big)$, where $\hat{\mathbf{M}} = (\hat{\mathbf{m}}_1,\ldots,\hat{\mathbf{m}}_N)\in\mathbb{R}^{N\times M}$ denotes the matrix of received signals with each row $\hat{\mathbf{m}}_i$ corresponding to a vector of symbols representing the codeword chosen by agent $i$, and likewise for $\mathbf{M} = (\mathbf{m}_1, \ldots, \mathbf{m}_N)\in\mathbb{R}^{N\times M}$ is the matrix of transmitted signals.
That is, the received signal of agent $i$ over the communication channel is a random function of the signals transmitted by all other agents, characterized by the conditional distribution of the multi-user communication channel.
In our simulations, we will consider independent and identically distributed channels as well as a channel with Markov noise, but our formulation is general enough to take into account arbitrarily correlated channels, both across time and users.

We can define a new Markov game with noisy communications, where the actions of agent $i$ now consist of two components, the environment actions $a_i^{(t)}$ as before, and the signal to be transmitted over the channel $\mathbf{m}_i^{(t)}$. 
Each agent, in addition to taking actions that affect the state of the environment, can also send signals to other agents over $M$ uses of the noisy communication channel. 
The observation of each agent is now given by $(o_i^{(t)}, \hat{\mathbf{m}}_i^{(t)})$; that is, a combination of the partial observation of the environment as before and the channel output signal.



At each time step $t$, agent $i$ observes $(o_i^{(t)}, \hat{\mathbf{m}}_i^{(t)})$ and selects an action $(a_i^{(t)}, \mathbf{m}_i^{(t)})$ according to its policy $\pi_i:\mathcal{O}_i \times \mathcal{C}_r^M \rightarrow \mathcal{A}_i \times \mathcal{C}_t^M$. 
The overall policy over all agents can be defined as $\Pi:\mathcal{S}\rightarrow\mathcal{A}$.
The objective of the Markov game with noisy communications is to maximize the discounted sum of rewards 
\begin{equation}
    V_\Pi(s)=\mathbb{E}_\Pi\Bigg[\sum_{t=1}^\infty\gamma^{t-1}r^{(t)}\Bigg|s^{(1)}=s\Bigg]
    \label{eq:value_function}
\end{equation}
for any initial state $s\in\mathcal{S}$ and $\gamma$ is the discount factor to ensure convergence.
We also define the state-action value function, also referred to as Q-function as 
\begin{equation}
    Q_\Pi(s^{(t)},a^{(t)})=\mathbb{E}_\Pi\Bigg[\sum_{i=t}^\infty\gamma^{(i-t)}r^{(t)}\Bigg|s^{(t)},a^{(t)}\Bigg].
    \label{eq:q_function}
\end{equation}


In the subsequent sections we will show that this formulation of the MA-POMDP with noisy communications lends itself to multiple problem domains where communication is vital to achieve non-trivial total reward values, and we devise methods that jointly learn to collaborate and communicate despite the noise in the channel. 
Although the introduced MA-POMDP framework with communications is fairly general and can model any multi-agent scenario with complex multi-user communications, our focus in this paper will be on point-to-point communications. 
This will allow us to expose the benefits of the joint communication and learning design, without having to deal with the challenges of multi-user communications. 
Extensions of the proposed framework to scenarios that would involve multi-user communication channels will be studied in future work. 

\section{Guided Robot with Point-to-Point Communications}
\label{subsec:eg_prob_guide_scout}

\begin{figure}
    \centering
    \includegraphics[width=0.3\linewidth]{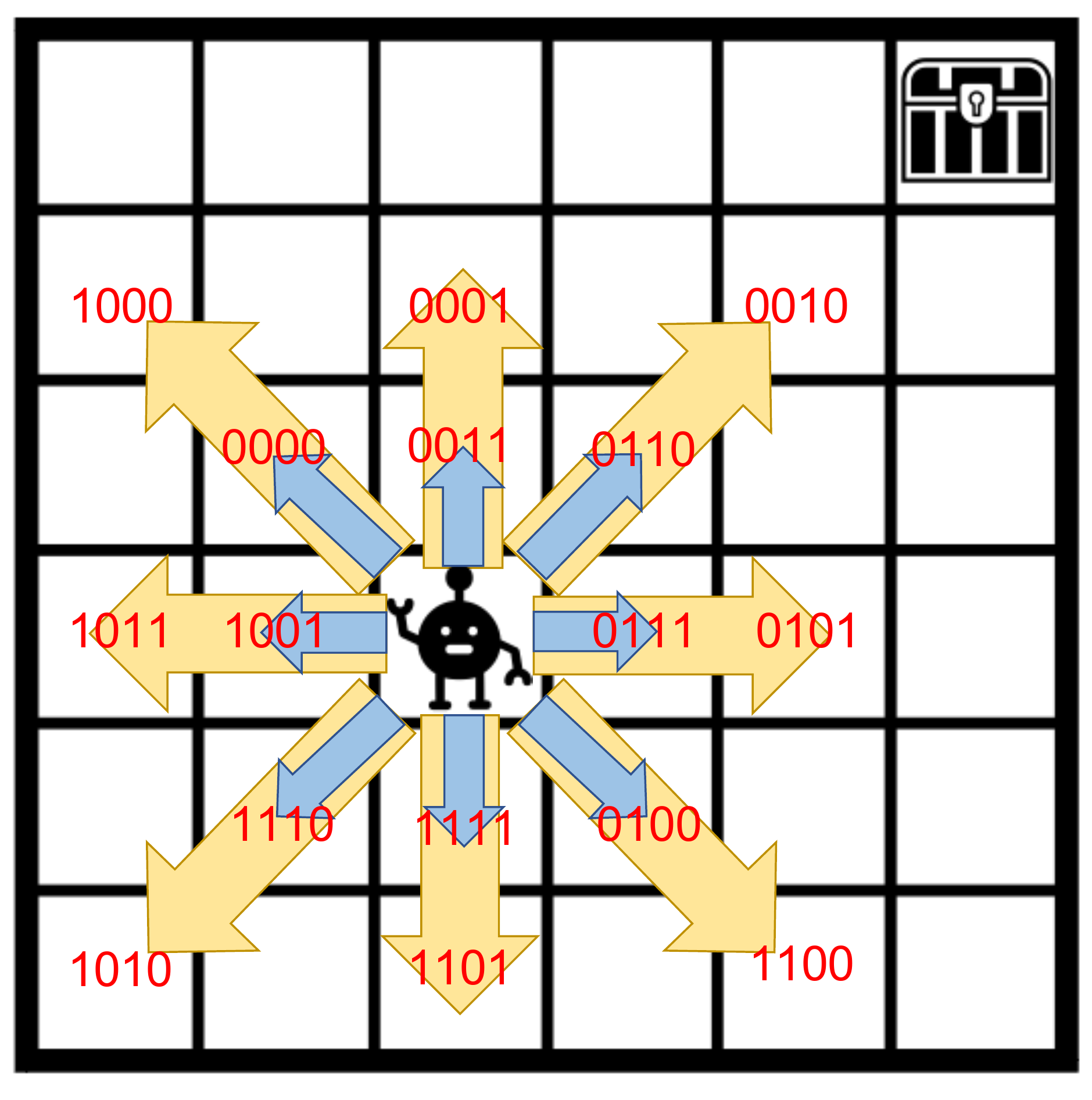}
    \caption{Illustration of the guided robot problem in grid world. The set $\mathcal{A}_2$ of 16 possible actions the scout agent can take using hand crafted (HC) codewords.}
    \label{fig:grid_world}
\vspace{-0.8cm}
\end{figure}

In this section, we consider a single-agent MDP and turn it into a MA-POMDP problem by dividing the single agent into two separate agents, a \textit{guide} and a \textit{scout}, which are connected through a noisy communication channel.
In this formulation, we assume that the guide observes the state of the original MDP perfectly, but cannot take actions on the environment directly. 
Contrarily, the scout can take actions on the environment, but cannot observe the environment state. 
Therefore, the guide communicates to the scout through a noisy communication channel and the scout has to take actions based on the signals it receives from the guide through the communication channel. 
The scout can be considered as a robot remotely controlled by the guide agent, which has sensors to observe the environment.

We consider this particular setting since it clearly exposes the importance of communication as the scout depends solely on the signals received from the guide. 
Without the communication channel, the scout is limited to purely random actions independent of the current state. 
Moreover, this scenario also allows us to quantify the impact of the channel noise on the overall performance since we recover the original single-agent MDP when the communication channel is perfect; that is, if any desired message can be conveyed over the channel in a reliable manner. 
Therefore, if the optimal reward for the original MDP can be determined, this would serve as an upper bound on the reward of the MA-POMDP with noisy communications. 

As an example to study the proposed framework and to develop and test numerical algorithms aiming to solve the obtained MA-POMDP problem, we consider a grid world of size $L\times L$, denoted by $\mathcal{L}= [L]\times[L]$, where $[L]=\{0,1,\dots,L-1\}$. We denote the scout position at time step $t$ by $p_s^{(t)}=(x_s^{(t)},y_s^{(t)})\in\mathcal{L}$. 
At each time instant, the scout can take one action from the set of 16 possible actions $\mathcal{A}=\{[1,0],[-1,0],[0,1],[0,-1],[1,1],[-1,1],[-1,-1],[1,-1],[2,0],$ $[-2,0],[0,2],[0,-2],[2,2],[-2,2],[-2,-2],[2,-2]\}$. See Fig. \ref{fig:awgn_grid_world} for an illustration of the scout and the 16 actions it can take. If the action taken by the scout ends up in a cell outside the grid world, the agent remains in its original location. 
The transition probability kernel of this MDP is specified as follows: after each action, the agent moves to the intended target location with probability (w.p.) $1-\delta$, and to a random neighboring cell w.p. $\delta$. 
That is, the next state is given by $s^{(t+1)} = s^{(t)} + a^{(t)}$ w.p. $1-\delta$, and $s^{(t+1)} = s^{(t)} + a^{(t)} + z^{(t)}$, where $z^{(t)}$ is uniformly distributed over the set $\{[1,0],[1,1], [0,1], [-1,1], [-1,0],[0,-1],[-1,-1],[1,-1] \}$ w.p. $\delta$. 

The objective of the scout is to find the treasure, located at $p_g=(x_g,y_g)\in\mathcal{L}$ as quickly as possible. 
We assume that the initial position of the scout and the location of the treasure are random, and are not the same. 
The scout takes instructions from the guide, who observes the grid world, and utilizes a noisy communication channel $M$ times to transmit signal $\mathbf{m}^{(t)}$ to the scout, who observes $\hat{\mathbf{m}}^{(t)}$ from the output of the channel.
To put it in the context of the MA-POMDP defined in Section \ref{sec:problem_formulation}, agent 1 is the guide, with observable state $o_1^{(t)} = s^{(t)}$, where $s^{(t)}=(p_s^{(t)},p_g)$, and action set $\mathcal{A}_1=\mathcal{C}_t$. 
Agent 2 is the scout, with observation $o_2^{(t)} = \hat{\mathbf{m}}^{(t)}$ and action set $\mathcal{A}_2 = \mathcal{A}$ (or, more precisely, $o_1^{(t)} = (s^{(t)},\o), o_2^{(t)} = (\o,\hat{\mathbf{m}}_2^{(t)})$). 
We define the reward function as follows to encourage the agents to collaborate to find the treasure as quickly as possible:
\begin{equation}
    r^{(t)}=\begin{cases}
                10,~&\text{if } p_s^{(t)}=p_g,\\
                -1,~&\text{otherwise}.
            \end{cases}
\end{equation}
The game terminates when $p_s^{(t)}=p_g$.

We should highlight that despite the simplicity of the problem, the original MDP is not a trivial one when both the initial state of the agent and the target location are random, as it has a rather large state space, and learning the optimal policy requires a long training process in order to observe all possible agent and target location pairs sufficiently many times. In order to simplify the learning of the optimal policy, and focus on learning the communication scheme, we will pay special attention to the scenario where $\delta=0$. 
This corresponds to the scenario in which the underlying MDP is deterministic, and it is not difficult to see that the optimal solution to this MDP is to take the shortest path to the treasure.


\begin{figure}
    \centering
    \includegraphics[width=0.6\linewidth]{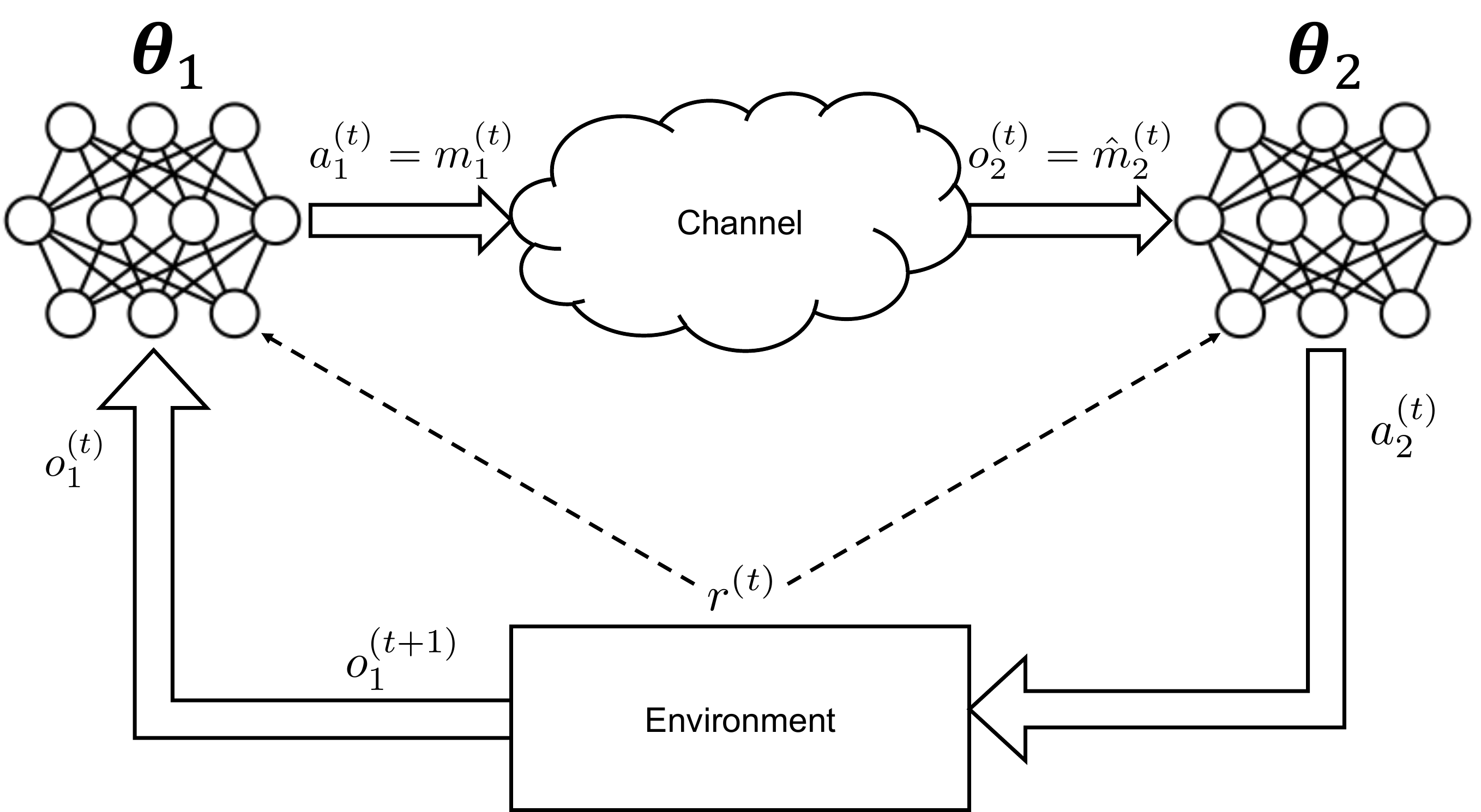}
    \caption{Information flow between the guide and the scout.}
    \label{fig:framework_grid_world}
\vspace{-0.8cm}
\end{figure}

We consider three types of channel distributions: the BSC, the AWGN, and the BN channel. 
In the BSC case, we have $\mathcal{C}_t = \{-1, +1\}$. 
For the AWGN channel and the BN channel, we have $\mathcal{C}_t = \{-1, +1\}$ if the input is constrained to binary phase shift keying (BPSK) modulation, or $\mathcal{C}_t =\mathbb{R}$ if no limitation is imposed on the input constellation. We will impose an average power constraint in the latter case. In both cases, the output alphabet is $\mathcal{C}_r = \mathbb{R}$. For the BSC, the output of the channel is given by $\hat{\mathbf{m}}_i^{(t)}=\mathbf{m}_i^{(t)} \oplus \mathbf{n}^{(t)}$, where $\mathbf{n}^{(t)} \sim \mathrm{Bernoulli(p_e)}$. 
For the AWGN channel, the output at the $i$th use of the channel is given by  $\hat{\mathbf{m}}_i^{(t)}=\mathbf{m}_i^{(t)}+\mathbf{n}^{(t)}$, where $\mathbf{n}^{(t)} \sim\mathcal{N}(0, \mathbf{I}_M\sigma_n^2)$ is the zero-mean Gaussian noise term with covariance matrix $\mathbf{I}_M\sigma_n^2$ and $\mathbf{I}_M$ is $M$-dimensional the identity matrix. 
For the BN channel, the output at the $i$th use of the channel is given by  $\hat{\mathbf{m}}_i^{(t)}=\mathbf{m}_i^{(t)}+\mathbf{n}_b^{(t)}$, where $\mathbf{n}_b^{(t)}$ is a two state Markov noise, with one state being the low noise state $N(0,\mathbf{I}_M\sigma_n^2)$ as in the AWGN case, and the other being the high noise state $N(0,\mathbf{I}_M(\sigma_n^2+\sigma_b^2))$. 
The probability of transitioning from the low noise state to the high noise state and remaining in that state is $p_b$. 
In practice, this channel models an occasional random interference from a nearby transmitter.

We first consider the BSC case, also studied in \cite{Roig:Globecom:20}. The action set of agent 1 is $\mathcal{A}_1=\{-1,+1\}^M$, while the observation set of agent 2 is $\mathcal{O}_2=\{-1,+1\}^M$. We will employ deep Q-learning network, introduced in \cite{mnih_human-level_2015}, which uses deep neural networks (DNNs) to approximate the Q-function in Eqn. (\ref{eq:q_function}).
More specifically, we use two distinct DNNs, parameterized by $\boldsymbol{\theta}_1$ and $\boldsymbol{\theta}_2$, respectively, representing DNNs for approximating the Q-functions of agent 1 (guide) and agent 2 (scout).
The guide observes $o_1^{(t)}=(p_s^{(t)}, p_g)$ and chooses a channel input signal $\mathbf{m}_1^{(t)}=a_1^{(t)}=\argmax_aQ_{\boldsymbol{\theta}_1}(o_1^{(t)},a)\in\mathcal{A}_1$, based on the current Q-function approximation. 
The signal is then transmitted across $M$ uses of the BSC. The scout observes $o_2^{(t)}=\hat{\mathbf{m}}_2^{(t)}$ at the output of the BSC, and chooses an action based on the current Q-function approximation $a_2^{(t)}=\argmax_a Q_{\boldsymbol{\theta}_2}(o_2^{(t)},a) \in \mathcal{A}_2$.
The scout then takes the action $a_2^{(t)}$, which updates its position $p_s^{(t+1)}$, collects reward $r^{(t)}$, and the process is repeated.
The reward $r^{(t)}$ is fed to both the guide and the scout to update $\boldsymbol{\theta}_1$ and $\boldsymbol{\theta}_2$.

As is typical in Q-learning methods, we use \textit{replay buffer}, \textit{target networks} and $\epsilon$-\textit{greedy} to improve the learned policy.
The replay buffers $\mathcal{R}_1$ and $\mathcal{R}_2$ store experiences $(o_1^{(t)},a_1^{(t)},r^{(t)},o_1^{(t+1)})$ and $(o_2^{(t)},a_2^{(t)},r^{(t)},o_2^{(t+1)})$ for the guide and scout, respectively, and we sample them uniformly to update the parameters $\boldsymbol{\theta}_1$ and $\boldsymbol{\theta}_2$.
This prevents the states from being correlated. 
We use target parameters ${\boldsymbol{\theta}_1^-}$ and ${\boldsymbol{\theta}_2^-}$, which are copies of ${\boldsymbol{\theta}_1}$ and ${\boldsymbol{\theta}_2}$, to compute the DQN loss function:
\begin{align}
    L_{\text{DQN}}(\boldsymbol{\theta}_i)=\frac{1}{2}\Big(r^{(t)}+\gamma\max_{a}\big\{Q_{\boldsymbol{\theta}_i^-}\big(o_i^{(t+1)},a\big)\big\} - Q_{\boldsymbol{\theta}_i} \big(o_i^{(t)},a_i^{(t)}\big)\Big)^2,~i=1,2.
    \label{eq:dqn_loss}
\end{align}
The parameters $\boldsymbol{\theta}_i$ are then updated via gradient descent according to the gradient $\nabla_{\boldsymbol{\theta}_i}L_{\text{DQN}}(\boldsymbol{\theta}_i)$, and the target network parameters are updated via
\begin{equation}
    \boldsymbol{\theta}_i^-\leftarrow\tau\boldsymbol{\theta}_i+(1-\tau)\boldsymbol{\theta}_i^-,~~i=1,2,
    \label{eq:target_update}
\end{equation}
where $0\leq\tau\leq1$.
Due to Q-learning being bootstrapped, if the same $Q_{\boldsymbol{\theta}_i}$ is used to estimate the state-action value of time step $t$ and $t+1$, both values would move at the same time, which may lead to the updates to never converge (like a dog chasing its tail).
By introducing the target networks, this effect is reduced due to the much slower updates of the target network, as done in Eqn. (\ref{eq:target_update}).

To promote exploration, we use $\epsilon$-greedy, which chooses a random action w.p. $\epsilon$ at each time step: 
\begin{equation}
    a_i^{(t)}=\begin{cases}
               \argmax_{a}Q_{\boldsymbol{\theta}_i}(o_i^{(t)},a),~&\text{w.p. }1-\epsilon\\
               a\sim\text{Uniform}(\mathcal{A}_i),~&\text{w.p. }\epsilon,
            \end{cases}
\end{equation}
where $a\sim\text{Uniform}(\mathcal{A}_i)$ denotes an action that is sampled uniformly from the action set $\mathcal{A}_i$.
The proposed solution for the BSC case is shown in Algorithm \ref{alg:robot_bsc}.

\begin{algorithm}[t]
\begin{small}
\SetAlgoLined
 Initialize Q networks, $\boldsymbol{\theta}_i,i=1,2$, using Gaussian $\mathcal{N}(0,10^{-2})$. Copy parameters to target networks $\boldsymbol{\theta}_i^-\leftarrow\boldsymbol{\theta}_i$.\\
 $\textit{episode}=0$\\
 \While{$\text{episode}<\text{episode-max}$}{
    $episode = episode + 1$\\
    $t=0$\\
    $\epsilon=\epsilon_{\text{end}}+(\epsilon_0-\epsilon_{\text{end}})e^{\big(\frac{\text{episode}}{-\lambda}\big)}$\\
     \While{Treasure NOT found OR $t<t_{\text{max}}$}{
        $t = t + 1$\\
        Observe $o_1^{(t)}=(p_s^{(t)},p_g)$\\
        $m_1^{(t)}=a_1^{(t)}
        =\begin{cases}
        \argmax_aQ_{\boldsymbol{\theta}_1}(o_1^{(t)},a),~\text{w.p. }1-\epsilon,\\
        a\sim\text{Uniform}(\mathcal{A}_1),~\text{w.p. }\epsilon.
        \end{cases}$\\
        Observe $o_2^{(t)}=P_{\text{BSC}}(\hat{m}_2^{(t)}|m_1^{(t)})$\\ 
        $a_2^{(t)}=\begin{cases}
        \argmax_aQ_{\boldsymbol{\theta}_1}(o_2^{(t)},a),~\text{w.p. }1-\epsilon,\\
        a\sim\text{Uniform}(\mathcal{A}_2),~\text{w.p. }\epsilon.
        \end{cases}$\\
        Take action $a_2^{(t)}$, collect reward $r^{(t)}$\\
        \uIf{$t>1$}{
            Store experiences:\\
            $(o_1^{(t-1)},a_1^{(t-1)},r^{(t-1)},o_1^{(t)})\in\mathcal{R}_1$ and  $(o_2^{(t-1)},a_2^{(t-1)},r^{(t-1)},o_2^{(t)})\in\mathcal{R}_2$
        }
     }
    Get batches $\mathcal{B}_1\subset\mathcal{R}_1$, $\mathcal{B}_2\subset\mathcal{R}_2$\\
    Compute DQN average loss $L_{\text{DQN}}(\boldsymbol{\theta}_i), i=1,2$ as in Eqn. (\ref{eq:dqn_loss}) using batch $\mathcal{B}_i$\\
    Update $\boldsymbol{\theta}_i$ using $\nabla_{\boldsymbol{\theta}_i}L_{\text{DQN}}(\boldsymbol{\theta}_i), i=1,2$.
    Update target networks $\boldsymbol{\theta}_i^-,i=1,2$ via Eqn. (\ref{eq:target_update})
 }
\caption{Proposed solution for the guided robot problem with BSC.}
\label{alg:robot_bsc}
\end{small}
\end{algorithm}

For the binary input AWGN and BN channels, we can use the exact same solution as the one used for BSC.
Note that the observation set of the scout is $\mathcal{O}_2=\mathbb{R}^M$.
However, the more interesting case is when $\mathcal{A}_1\in\mathbb{R}^M$.
It has been observed in the JSCC literature \cite{tung_sparsecast:_2018,bourtsoulatze_deep_2018}, that relaxing the constellation constraints, similar to analog communications, and training the JSCC scheme in an end-to-end fashion can provide significant performance improvements thanks to the greater degree of freedom available to the transmitter.
In this case, since the guide can output continuous actions, we can employ the deep deterministic policy gradient (DDPG) algorithm proposed in \cite{lillicrap_continuous_2019}.
DDPG uses a parameterized policy function $\mu_{\boldsymbol{\psi}}(o_1^{(t)})$, which specifies the current policy by deterministically mapping the observation $o_1^{(t)}$ to a continuous action.
The critic $Q_{\boldsymbol{\theta}_1}(o_1^{(t)},\mu_{\boldsymbol{\psi}}(o_1^{(t)}))$, then estimates the value of the action taken by $\mu_{\boldsymbol{\psi}}(o_1^{(t)})$, and is updated as it is with DQN in Eqn. (\ref{eq:dqn_loss}).

The guide policy is updated by applying the chain rule to the expected return from the initial distribution 
\begin{align}
    J=\mathbb{E}_{o_1^{(t)}\sim\rho^{\pi_1},o_2^{(t)}\sim\rho^{\pi_2},a_1^{(t)}\sim\pi_1,a_2^{(t)}\sim\pi_2}\Bigg[\sum_{t=1}^\infty\gamma^{t-1}r^{(t)}(o_1^{(t)},o_2^{(t)},a_1^{(t)},a_2^{(t)})\Bigg],
    \label{eq:exp_return}
\end{align}
where $\rho^{\pi_i}$ is the discounted observation visitation distribution for policy $\pi_i$.
Since we solve this problem by letting each agent treat the other agent as part of the environment, the value of the action taken by the guide is only dependent on its observation $o_1^{(t)}$ and action $\mu_{\boldsymbol{\psi}}(o_1^{(t)})$.
Thus, we use a result in \cite{silver_deterministic_2014} where the gradient of the objective $J$ in Eqn. (\ref{eq:exp_return}) with respect to the guide policy parameters $\boldsymbol{\psi}$ is shown to be
\begin{align}
    \nabla_{\boldsymbol{\psi}} J &=\mathbb{E}_{o_1^{(t)}\sim\rho^{\pi_1}}\Big[\nabla_{\boldsymbol{\psi}} Q_{\boldsymbol{\theta}_1}(o,a)\big|_{o=o_1^{(t)},a=\mu_{\boldsymbol{\psi}}(o_1^{(t)})}\Big]\\
    &=\mathbb{E}_{o_1^{(t)}\sim\rho^{\pi_1}}\Big[\nabla_a Q_{\boldsymbol{\theta}_1}(o,a)\big|_{o=o_1^{(t)},a=\mu_{\boldsymbol{\psi}}(o_1^{(t)})}\nabla_{\boldsymbol{\psi}}\mu_{\boldsymbol{\psi}}(o)\big|_{o=o_1^{(t)}}\Big]
    \label{eq:ddpg_gradient}
\end{align}
 if certain conditions specified in Theorem \ref{thm:ddpg_compatibility} are satisfied.
\begin{theorem}[{{\cite{silver_deterministic_2014}}}]
    A function approximator $Q_{\boldsymbol{\theta}}(o,a)$ is compatible (i.e., the gradient of the true Q function $Q_{\boldsymbol{\theta}^\ast}$ is preserved by the function approximator) with a deterministic policy $\mu_{\boldsymbol{\psi}}(o)$, such that $\nabla_{\boldsymbol{\psi}} J(\boldsymbol{\psi})=\mathbb{E}[\nabla_{\boldsymbol{\psi}}\mu_{\boldsymbol{\psi}}(o)\nabla_aQ_{\boldsymbol{\theta}}(o,a)|_{a=\mu_{\boldsymbol{\psi}}(o)}]$, if 
    \begin{enumerate}
        \item $\nabla_aQ_{\boldsymbol{\theta}}(o,a)|_{a=\mu_{\boldsymbol{\psi}}(o)}=\nabla_{\boldsymbol{\psi}}\mu_{\boldsymbol{\psi}}(o)^\top\boldsymbol{\theta}$, and 
        \item $\boldsymbol{\theta}$ minimizes the mean-squared error,
        $\mathbb{E}[e(o;\boldsymbol{\theta},\boldsymbol{\psi})^\top e(o;\boldsymbol{\theta},\boldsymbol{\psi})]$, where\\
        $e(o;\boldsymbol{\theta},\boldsymbol{\psi})\!=\!\nabla_a\big[Q_{\boldsymbol{\theta}}(o,a)|_{a=\mu_{\boldsymbol{\psi}}(o)}-Q_{\boldsymbol{\theta}^\ast}(o,a)|_{a=\mu_{\boldsymbol{\psi}}(o)}\big]$,\\
        and $\boldsymbol{\theta}^\ast$ are the parameters that describe the true Q function exactly.
    \end{enumerate}
\label{thm:ddpg_compatibility}
\end{theorem}
In practice, criterion 2) of Theorem \ref{thm:ddpg_compatibility} is approximately satisfied via mean-squared error loss and gradient descent, but criterion 1) may not be satisfied.
Nevertheless, DDPG works well in practice.

The DDPG loss is two-fold: the critic loss is computed as 
\begin{align} \label{eq:ddpg_critic_loss}
    L_{\text{DDPG}}^{\text{Critic}}(\boldsymbol{\theta}_1)=\Big(r^{(t)}+\gamma\Big\{Q_{\boldsymbol{\theta}_1^-}(o_1^{(t+1)},\mu_{\boldsymbol{\psi}^-}(o_1^{(t+1)}))\Big\} - Q_{\boldsymbol{\theta}_1}(o_1^{(t)},\mu_{\boldsymbol{\psi}}(o_1^{(t)})\Big)^2,
\end{align}
whereas the policy loss is computed as
\begin{align}
    &L_{\text{DDPG}}^{\text{Policy}}(\psi)=-Q_{\boldsymbol{\theta}_1}(o_1^{(t)},\mu_{\boldsymbol{\psi}}(o_1^{(t)})).\label{eq:ddpg_policy_loss}
\end{align}

As with the DQN case, we can also use a replay buffer and target network to train the DDPG policy. To promote exploration, we add noise to the actions taken as follows:
\begin{equation}
    a_1^{(t)}=\mu_{\boldsymbol{\psi}}(o_1^{(t)}) + w^{(t)},
\end{equation}
where $w^{(t)}$ is an Orstein-Uhlenbeck process \cite{uhlenbeck_theory_1930} to generate temporally correlated noise terms. The proposed solution for the AWGN and BN channel is summarized in Algorithm \ref{alg:robot_awgn}. We find that by relaxing the modulation constraint to $\mathbb{R}^M$, the learned policies of guide and scout are substantially better than those achieved in the BPSK case. The numerical results illustrating this conclusion will be discussed in Section \ref{sec:results}.

\begin{algorithm}[]
\begin{small}
\caption{Proposed solution for guided robot problem for AWGN and BN channel.}\label{alg:robot_awgn}
\SetAlgoLined
 Initialize Q networks $\boldsymbol{\theta}_i,i=1,2$, using Gaussian $\mathcal{N}(0,10^{-2})$ and policy network $\boldsymbol{\psi}$ if $\mathcal{A}_1\in\mathbb{R}^M$.
 Copy parameters to target networks $\boldsymbol{\theta}_i^-\leftarrow\boldsymbol{\theta}_i$, $\boldsymbol{\psi}^-\leftarrow\boldsymbol{\psi}$.\\
 $\textit{episode}=1$\\
 \While{$\text{episode}<\text{episode-max}$}{
    $t=1$\\
    $\epsilon=\epsilon_{\text{end}}+(\epsilon_0-\epsilon_{\text{end}})e^{\big(\frac{\text{episode}}{-\lambda}\big)}$\\
     \While{Treasure NOT found OR $t<t_{\text{max}}$}{
        Observe $o_1^{(t)}=(p_s^{(t)},p_g)$\\
        \uIf{$\mathcal{A}_1=\{-1,+1\}^M$}{
            $m_1^{(t)}=a_1^{(t)}  =\begin{cases}
            \argmax_aQ_{\boldsymbol{\theta}_1}(o_1^{(t)},a),~\text{w.p. }1-\epsilon,\\
            a\sim\text{Uniform}(\mathcal{A}_1),~\text{w.p. }\epsilon.
            \end{cases}$\\
        }
        \uElseIf{$\mathcal{A}_1=\mathbb{R}^M$}{
            $m_1^{(t)}=\mu_\psi(o_1^{(t)})+w^{(t)}$\\
            Normalize $m_1^{(t)}$ via Eqn. (\ref{eq:power_norm})
        }
        Observe $o_2^{(t)}=P_{\text{AWGN}}(\hat{m}_2^{(t)}|m_1^{(t)})$ or $P_{\text{BN}}(\hat{m}_2^{(t)}|m_1^{(t)})$\\ 
        $a_2^{(t)}=\begin{cases}
        \argmax_aQ_{\boldsymbol{\theta}_1}(o_2^{(t)},a),~\text{w.p. }1-\epsilon,\\
        a\sim\text{Uniform}(\mathcal{A}_2),~\text{w.p. }\epsilon.
        \end{cases}$\\
        Take action $a_2^{(t)}$, collect reward $r^{(t)}$\\
        \uIf{$t>1$}{
            Store experiences:\\
            $(o_1^{(t-1)},a_1^{(t-1)},r^{(t-1)},o_1^{(t)})\in\mathcal{R}_1$ \mbox{ and }            $(o_2^{(t-1)},a_2^{(t-1)},r^{(t-1)},o_2^{(t)})\in\mathcal{R}_2$
        }
         $t=t+1$
     }
    Compute average scout loss $L_{\text{DQN}}(\boldsymbol{\theta}_2)$ as in Eqn. (\ref{eq:dqn_loss}) using batch $\mathcal{B}_2 \subset \mathcal{R}_2$\\
    Update $\boldsymbol{\theta}_2$ using $\nabla_{\boldsymbol{\theta}_2}L_{\text{DQN}}(\boldsymbol{\theta}_2)$\\
    \uIf{$\mathcal{A}_1=\{-1,+1\}^M$}{
        Compute DQN average loss $L_{\text{DQN}}(\boldsymbol{\theta}_1)$ as in Eqn. (\ref{eq:dqn_loss}) using batch $\mathcal{B}_1 \subset \mathcal{R}_1$\\
        Update $\boldsymbol{\theta}_1$ using $\nabla_{\boldsymbol{\theta}_1}L_{\text{DQN}}(\boldsymbol{\theta}_1)$\\
        Update target network $\boldsymbol{\theta}_i^-,i=1,2$ via Eqn. (\ref{eq:target_update})
    }
    \uElseIf{$\mathcal{A}_1=\mathbb{R}^M$}{
        Compute average DDPG Critic loss $L_{\text{DDPG}}^{\text{Critic}}(\boldsymbol{\theta}_1)$ as in Eqn. (\ref{eq:ddpg_critic_loss}) using batch $\mathcal{B}_1$\\
        Compute average DDPG Policy loss $L_{\text{DDPG}}^{\text{Policy}}(\boldsymbol{\psi})$ as in Eqn. (\ref{eq:ddpg_policy_loss}) using batch $\mathcal{B}_1$\\
        Update $\boldsymbol{\theta}_1$ and $\boldsymbol{\psi}$ using $\nabla_{\boldsymbol{\theta}_1}L_{\text{DDPG}}^{\text{Critic}}(\boldsymbol{\theta}_1)$ and $\nabla_{\psi}L_{\text{DDPG}}^{\text{Policy}}(\boldsymbol{\psi})$\\
        Update target network $\boldsymbol{\theta}_i^-,i=1,2,\boldsymbol{\psi}^-$ via Eqn. (\ref{eq:target_update})
    }
    $\text{episode}=\text{episode}+1$
 }
\end{small}
\end{algorithm}

To ensure that the actions taken by the guide meet the power constraint we normalize the channel input to an average power of $1$ as follows:
\begin{equation}
    a_1^{(t)}[k]\leftarrow\sqrt{M}\frac{a_1^{(t)}[k]}{\sqrt{\Big(a_1^{(t)}\Big)^\top a_1^{(t)}}},~k=1,\dots,M.
    \label{eq:power_norm}
\end{equation}
The signal-to-noise ratio (SNR) of the AWGN channel is then defined as 
\begin{equation}
    \text{SNR}=-10\log_{10}(\sigma_n^2)~\text{(dB)}.
\end{equation}
Due to the burst noise, we define SNR of the BN channel by the expected SNR of the two noise states: 
\begin{equation}
    \text{SNR}=-10((1-p_b)\log_{10}(\sigma_n^2)+p_b\log_{10}(\sigma_n^2+\sigma_b^2))~\text{(dB)}.
\end{equation}

In Section \ref{sec:results}, we will study the effects of both the channel SNR and the channel bandwidth on the performance. Naturally, the capacity of the channel increases with both the SNR and the bandwidth. However, we would like to emphasize that the Shannon capacity is not a relevant metric \textit{per se} for the problem at hand. Indeed, we will observe that the benefits from increasing channel bandwidth and channel SNR saturate beyond some point. Nevertheless, the performance achieved for the underlying single-agent MDP assuming a perfect communication link from the guide to the scout serves as a more useful bound on the performance with any noisy communication channel. 
The numerical results for this example will be discussed in detail in Section \ref{sec:results}.

\section{Joint Channel Coding and Modulation}
\label{subsec:eg_prob_channel_coding}

The formulation given in Section \ref{sec:problem_formulation} can be readily extended to the aforementioned classic ``level A" communication problem of channel coding and modulation. 
Channel coding is a problem where $B$ bits are communicated over $M$ channel uses, which corresponds to a code rate of $B/M$ bits per channel use.
In the context of the Markov game introduced previously, we can consider $2^B$ states corresponding to each possible message. Agent 2 has $2^B$ actions, each corresponding to a different reconstruction of the message at agent 1. 
All the actions transition to the terminal state. 
The transmitter observes the state and sends a message by using the channel $M$ times, and the receiver observes a noisy version of the message at the output of the channel and chooses an action.
Herein, we consider the scenario with real channel input and output values, and an average power constraint on the transmitted signals at each time $t$. 
As such, we can define $\mathcal{O}_1=\mathcal{A}_2 = \{0,1\}^B$ and $\mathcal{A}_1 = \mathcal{O}_2 = \mathcal{C}^M_t$. We note that maximizing the average reward in this problem is equivalent to designing a channel code with blocklength $B$ and rate $B/M$ with minimum BLER. 

\begin{figure}
    \centering
    \includegraphics[width=.6\linewidth]{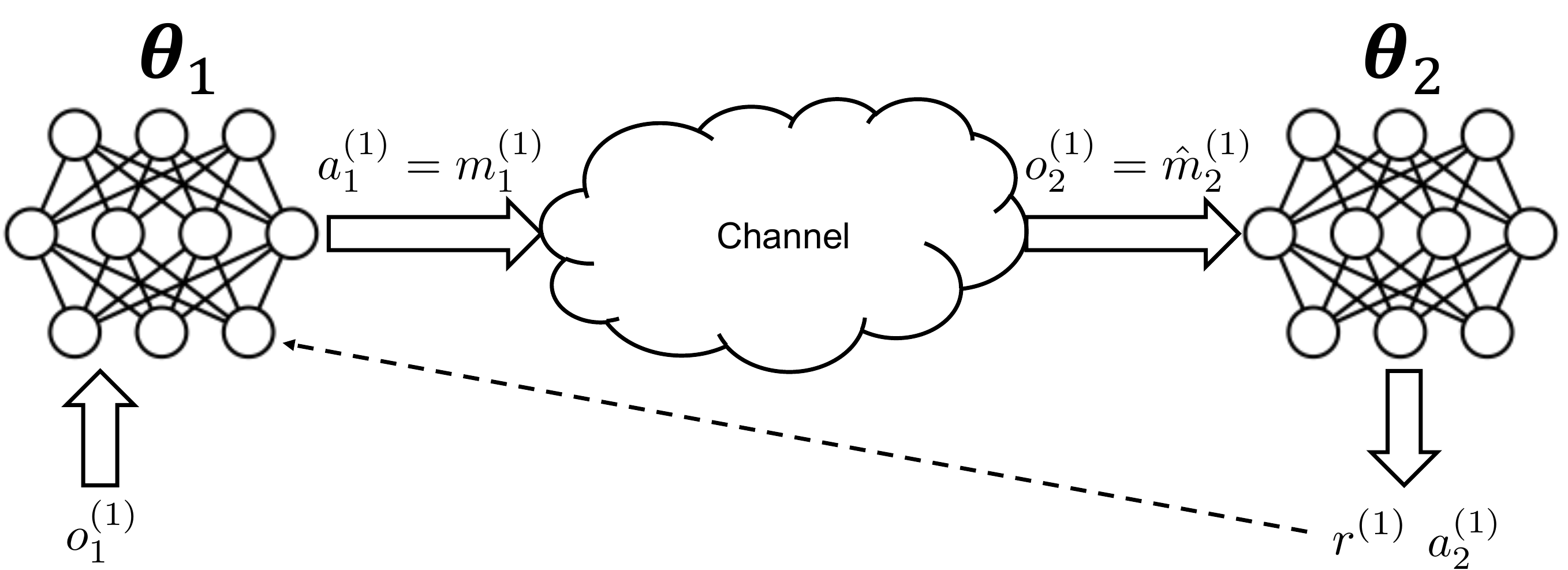}
    \caption{Information flow between the transmitter and the receiver.}
    \label{fig:framework_ch_coding}
\vspace{-0.5cm}
\end{figure}

There have been many recent studies focusing on the design of channel coding and modulation schemes using machine learning techniques \cite{Nachmani:STSP:18, Dorner:Asilomar:17, Felix:SPAWC:18, bourtsoulatze_deep_2018, Kurka:JSAIT:20, aoudia_model-free_2019}. Most of these works use supervised learning techniques, assuming a known and differentiable channel model, which allows backpropagation through the channel during training. On the other hand, here we assume that the channel model is not known, and the agents are limited to their observations of the noisy channel output signals, and must learn a communication strategy through trial and error.

A similar problem is considered in \cite{aoudia_model-free_2019} from a supervised learning perspective. The authors show that by approximating the gradient of the transmitter with the stochastic policy gradient of the vanilla REINFORCE algorithm \cite{williams_simple_1992}, it is possible to train both the transmitter and the receiver without knowledge of the channel model. We wish to show here that this problem is actually a special case of the problem formulation we constructed in Section \ref{sec:problem_formulation} and that by approaching this problem from a RL perspective, the problem lends itself to a variety of solutions from the vast RL literature.

\begin{algorithm}[t]
\begin{small}
\caption{Proposed solution for joint channel coding-modulation problem.}
\label{alg:channel_coding}
\SetAlgoLined
 Initialize DNNs $\boldsymbol{\theta}_i,i=1,2$, with Gaussian $\mathcal{N}(0,10^{-2})$, and policy network $\boldsymbol{\psi}$ if using DDPG.\\
 $\textit{episode}=1$\\
 \While{$\text{episode}<\text{episode-max}$}{
    $\epsilon=\epsilon_{\text{end}}+(\epsilon_0-\epsilon_{\text{end}})e^{-\frac{\text{episode}}{\lambda}}$\\
    Observe $o_1^{(1)}\sim\text{Uniform}(\mathcal{O}_1)$\\
    $m_1^{(1)}=\mu_{\boldsymbol{\psi}}(o_1^{(1)})+w^{(1)}$\\
    Normalize $m_1^{(1)}$ via Eqn. (\ref{eq:power_norm})\\
    Observe $o_2^{(1)}=P_{\text{AWGN}}(\hat{m}_2^{(1)}|m_1^{(1)})$ or $P_{\text{BN}}(\hat{m}_2^{(1)}|m_1^{(1)})$\\ 
    $a_2^{(1)}=\argmax_aQ_{\boldsymbol{\theta}_1}(o_2^{(1)},a)
    $\\
    Collect reward $r^{(1)}$ \\
    Store experiences:\\
    $(o_1^{(1)},a_1^{(1)},r^{(1)})\in\mathcal{R}_1$ and $(o_2^{(1)},a_2^{(1)},r^{(1)})\in\mathcal{R}_2$\\
    Get batches $\mathcal{B}_1\subset\mathcal{R}_1$, $\mathcal{B}_2\subset\mathcal{R}_2$\\
    Compute average receiver loss $L_{\text{CE}}(o_2^{(1)};\boldsymbol{\theta}_2)$ as in Eqn. (\ref{eq:ce_reward}) using batch $\mathcal{B}_2$\\
    Update $\boldsymbol{\theta}_2$ using $\nabla_{\boldsymbol{\theta}_2}L_{\text{CE}}(o_2^{(1)};\boldsymbol{\theta}_2)$\\
    \uIf{use DDPG}{
        Compute average transmitter losses $L_{\text{DDPG}}^{\text{Critic}}(\boldsymbol{\theta}_1)$ and $L_{\text{DDPG}}^{\text{Policy}}(\boldsymbol{\psi})$ as in Eqns. (\ref{eq:ch_coding_ddpg_critic_loss},\ref{eq:ch_coding_ddpg_policy_loss}) using $\mathcal{B}_1$\\
        Update $\boldsymbol{\theta}_1$ and $\boldsymbol{\psi}$ $\nabla_{\boldsymbol{\theta}_1}L_{\text{DDPG}}^{\text{Critic}}(\boldsymbol{\theta}_1)$ and $\nabla_{\boldsymbol{\psi}} L_{\text{DDPG}}^{\text{Policy}}(\boldsymbol{\psi})$
    }
    \uElseIf{use REINFORCE}{
        Compute average transmitter gradient $\nabla_{\boldsymbol{\theta}_1}J(\boldsymbol{\theta}_1)$ as in Eqn. (\ref{eq:reinforce_loss}) using $\mathcal{B}_1$\\
        Update $\boldsymbol{\theta}_1$ using $\nabla_{\boldsymbol{\theta}_1}J(\boldsymbol{\theta}_1)$
    }
    \uElseIf{use Actor-Critic}{
        Compute average transmitter loss $\nabla_{\boldsymbol{\theta}_1}J(\boldsymbol{\theta}_1)$ as in Eqn. (\ref{eq:a2c_loss}) using $\mathcal{B}_1$\\
        Update $\boldsymbol{\theta}_1$ using $\nabla_{\boldsymbol{\theta}_1}J(\boldsymbol{\theta}_1)$\\
        Update value estimate $v_{\pi_1}(o_1^{(1)})$ via Eqn. (\ref{eq:value_estimate})
    }
    $\text{episode}=\text{episode}+1$
 }
\end{small}
\end{algorithm}

Here, we opt to use DDPG to learn a deterministic joint channel coding-modulation scheme and use the DQN algorithm for the receiver, as opposed to the vanilla REINFORCE algorithm used in \cite{aoudia_model-free_2019}.
We use negative cross-entropy (CE) loss as the reward function:
\begin{equation}
    r^{(1)}=-L_{\text{CE}}(\hat{m}^{(1)}_1)=\sum_{k=1}^{2^B}\log(Pr(c_k|\hat{m}^{(1)}_1)),
    \label{eq:ce_reward}
\end{equation}
where $c_k$ is the $k$th codeword in $\mathcal{O}_1$.
The receiver DQN is trained simply with the CE loss, while the transmitter DDPG algorithm receives the reward $r^{(1)}$.
Similar to the \textit{guided robot} problem in Section \ref{subsec:eg_prob_guide_scout}, we use replay buffer to improve the training process.
We note here that in this problem, each episode is simply a one-step MDP, as there is no state transition.
As such, the replay buffers store only $(o_1^{(1)},a_1^{(1)},r^{(1)})$, $(o_2^{(1)},a_2^{(1)},r^{(1)})$ and a target network is not required.
Consequently, the DDPG losses can be simplified as
\begin{align}
    &L_{\text{DDPG}}^{\text{Critic}}(\boldsymbol{\theta}_1)=\Big(Q_{\boldsymbol{\theta}_1}(o_1^{(1)},\mu_{\boldsymbol{\psi}}(o_1^{(1)})-r^{(1)}\Big)^2,\label{eq:ch_coding_ddpg_critic_loss}\\
    &L(\boldsymbol{\psi})_{\text{DDPG}}^{\text{Policy}}=-Q_{\boldsymbol{\theta}_1}(o_1^{(1)},\mu_{\boldsymbol{\psi}}(o_1^{(1)}))\label{eq:ch_coding_ddpg_policy_loss}
\end{align}

Furthermore, we improve upon the algorithm used in \cite{aoudia_model-free_2019} by implementing a critic, which estimates the advantage of a given state-action pair by subtracting a baseline from policy gradient.
That is, in the REINFORCE algorithm, the gradient is estimated as
\begin{equation}
    \nabla_{\boldsymbol{\theta}_1} J(\boldsymbol{\theta}_1)=\nabla_{\boldsymbol{\theta}_1}\log\pi_1(a_1^{(1)}|o^{(1)}_1;\boldsymbol{\theta}_1)r^{(1)} \;.
    \label{eq:reinforce_loss}
\end{equation}
It is shown in \cite{konda_actor-critic_nodate} that by subtracting a baseline $b(o_1^{(1)})$, the variance of the gradient $\nabla_{\boldsymbol{\theta}} J(\boldsymbol{\theta})$ can be greatly reduced. 
Herein, we use the value of the state, defined by Eqn. (\ref{eq:value_function}), except, in this problem, the trajectories all have length 1.
Therefore, the value function can be simplified to 
\begin{equation}
    b(o_1^{(1)})=v_{\pi_1}(o_1^{(1)})=\mathbb{E}_{\pi_1}\big[r^{(1)}|o_1^{(1)}\big].
\label{eq:ch_code_baseline}
\end{equation}
The gradient of the policy with respect to the expected return $J(\boldsymbol{\theta}_1)$ is then 
\begin{equation}
    \nabla_{\boldsymbol{\theta}_1} J(\boldsymbol{\theta}_1)=\nabla_{\boldsymbol{\theta}_1}\log\pi_1(a_1^{(1)}|o_1^{(1)};\boldsymbol{\theta}_1)(r^{(1)}-v_{\pi_1}(o_1^{(1)})).
    \label{eq:a2c_loss}
\end{equation}
In practice, to estimate $v_{\Pi}(o^{(1)}_1)$, we use a weighted moving average of the reward collected for a given state $o_1^{(1)}\in\mathcal{O}_1$ in $\mathcal{B}_1(o_1^{(1)})=\{(o,a)\in \mathcal{B}_1| o=o_1^{(1)}\}$
for the batch of trajectories $\mathcal{B}_1$:
\begin{equation}
    v_{\pi_1}(o_1^{(1)})\leftarrow
    (1-\alpha) v_{\pi_1}(o_1^{(1)})+
    \frac{\alpha}{|\mathcal{B}_1(o_1^{(1)})|}\!\!\sum_{(o,a)\in \mathcal{B}_1(o_1^{(1)})}\!\! r^{(1)}(o,a),
\label{eq:value_estimate}
\end{equation}
where $\alpha$ is the weight of the average and $v_{\pi_1}(o_1^{(1)})$ is initialized with zeros.
We use $\alpha=0.01$ in our experiments.
The algorithm for solving the joint channel coding and modulation problem is shown in Algorithm \ref{alg:channel_coding}.
The numerical results and comparison with alternative designs are presented in the next section.





\section{Numerical Results}
\label{sec:results}


\begin{table}
\begin{center}
\caption{DNN architecture and hyperparameters used.}
\begin{tabular}{|c|c|c|}
\hline
$Q_{\boldsymbol{\theta}_i}$ & $\mu_{\boldsymbol{\psi}}$ & Hyperparameters \\ \hline
Linear: 64 & Linear: 64 & $\gamma=0.99$ \\
ReLU & ReLU & $\epsilon_0=0.9$ \\
Linear: 64 & Linear: 64 & $\epsilon_{\text{end}}=0.05$ \\
ReLU & ReLU & $\lambda=1000$ \\
Linear: $\begin{cases}
                |\mathcal{A}_i|,~&\text{if DQN}, \\ 
                1,~&\text{if DDPG}
        \end{cases}$
& Linear: dim$(\mathcal{A}_i)$ & $\tau=0.005$  \\ \hline
\end{tabular}
\label{tab:parameters}
\end{center}
\vspace{-0.8cm}
\end{table}

We first define the DNN architecture used for all the experiments in this section.
For networks, the inputs are processed by three fully connected layers followed by the rectified linear unit (ReLU) activation function.
The weights of the layers are initialized using Gaussian initialization with mean 0 and standard deviation $0.01$.
We store $100K$ experience samples in the replay buffer ($|\mathcal{R}_i|=100K$), and sample batches of size $128$ for training.
We train every experiment for $500K$ episodes.
The function used for $\epsilon$-greedy exploration is
\begin{equation}
    \epsilon=\epsilon_{\text{end}}+(\epsilon_0-\epsilon_{\text{end}})e^{\big(-\frac{\text{episode}}{\lambda}\big)}
\end{equation}
where $\lambda$ controls the decay rate of $\epsilon$.
We use the ADAM optimizer \cite{kingma_adam_2017} with learning rate $0.001$ for all the experiments.
The network architectures and the hyperparameters chosen are summarized in Table \ref{tab:parameters}.
We consider $\text{SNR}\in[0,23]$ dB for the AWGN channel. 
For the BN channel, we use the same SNR range as the AWGN channel for the low noise state and set $\sigma_b=2$ for the high noise state.
We consider $p_b\in\{0.1,0.2\}$ to see the effect of changing the high noise state probability.



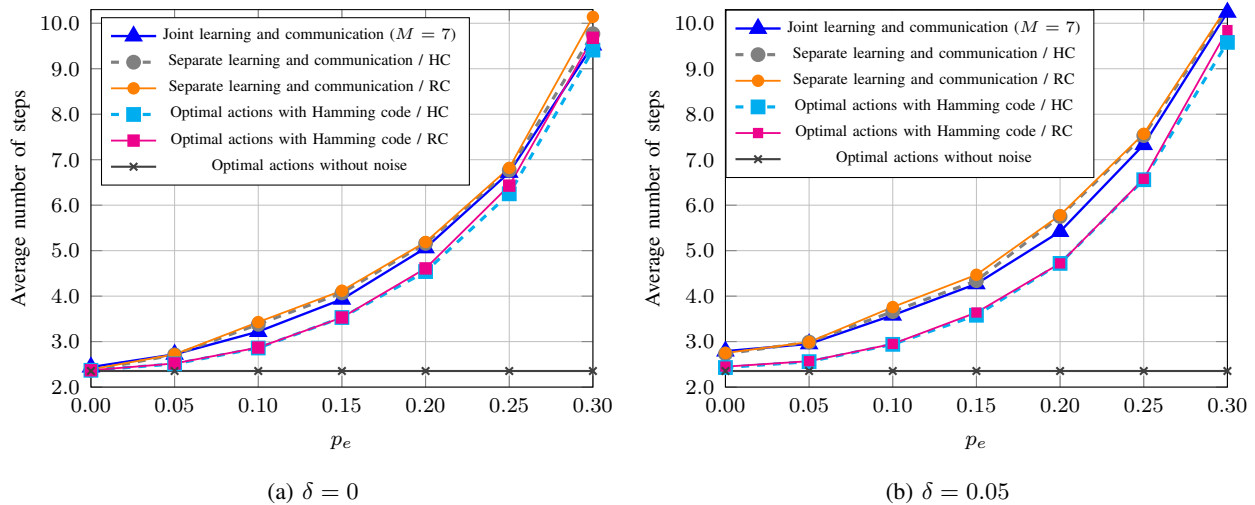
\begin{figure} 
    \centering
  \subfloat[$\delta=0$ \label{subfig:bsc_grid_world}]{%
    \begin{tikzpicture}
        \pgfplotsset{
            legend style={
                font=\fontsize{6}{6}\selectfont,
                at={(0.02,.98)},
                anchor=north west,
            },
            height=0.4\linewidth,
            width=0.5\linewidth,
            xmin=0,
            xmax=0.3,
            ymin=2.,
            ymax=10.3,
            ytick distance=1,
            xlabel={$p_e$},
            ylabel={Average number of steps},
            grid=both,
            grid style={line width=.1pt, draw=gray!10},
            major grid style={line width=.2pt,draw=gray!50},
            every axis/.append style={
                x label style={
                    font=\fontsize{8}{8}\selectfont,
                    at={(axis description cs:0.5,-0.1)},
                    },
                y label style={
                    font=\fontsize{8}{8}\selectfont,
                    at={(axis description cs:-0.1,0.5)},
                    },
                x tick label style={
                    font=\fontsize{8}{8}\selectfont,
                    /pgf/number format/.cd,
                    fixed,
                    fixed zerofill,
                    precision=2,
                    /tikz/.cd
                    },
                y tick label style={
                    font=\fontsize{8}{8}\selectfont,
                    /pgf/number format/.cd,
                    fixed,
                    fixed zerofill,
                    precision=1,
                    /tikz/.cd
                    },
            }
        }
        \begin{axis}
        \addplot[blue, solid, line width=0.9pt, mark=triangle*, mark options={fill=blue, scale=1.6}] 
                table [x=PE, y=BINARY, col sep=comma] {Data/bsc_grid_world.csv};
        \addlegendentry{Joint learning and communication ($M=7$)}
        
        \addplot[color=gray, dashed, line width=1.2pt, mark=*, mark options={fill=gray, solid, scale=1.1}] 
        table [x=PE, y=HAMMING (HC), col sep=comma] {Data/bsc_grid_world.csv};
        \addlegendentry{Separate learning and communication / HC}
        
        \addplot[color=orange, solid, line width=0.7pt, mark=*, mark options={fill=orange, scale=1}] 
        table [x=PE, y=HAMMING (RC), col sep=comma] {Data/bsc_grid_world.csv};
        \addlegendentry{Separate learning and communication / RC}
        
        \addplot[color=cyan, dashed, line width=1.2pt, mark=square*, mark options={fill=cyan, solid, scale=1.1}] 
        table [x=PE, y=OPT (HC), col sep=comma] {Data/bsc_grid_world.csv};
        \addlegendentry{Optimal actions with Hamming code / HC}
        
        \addplot[color=magenta, solid, line width=0.7pt, mark=square*, mark options={fill=magenta, scale=1}] 
        table [x=PE, y=OPT (RC), col sep=comma] {Data/bsc_grid_world.csv};
        \addlegendentry{Optimal actions with Hamming code / RC}
        
        \addplot[color=darkgray, solid, thick, mark=x, mark options={fill=darkgray, scale=1}] 
        table [x=PE, y=LB, col sep=comma] {Data/bsc_grid_world.csv};
        \addlegendentry{Optimal actions without noise}
        \end{axis}
        \end{tikzpicture}
    }
  \subfloat[$\delta=0.05$ \label{subfig:bsc_grid_world_noisy}]{%
    \begin{tikzpicture}
        \pgfplotsset{
            legend style={
                font=\fontsize{6}{6}\selectfont,
                at={(0.,1.)},
                anchor=north west,
            },
            height=0.4\linewidth,
            width=0.5\linewidth,
            xmin=0,
            xmax=0.3,
            ymin=2.,
            ymax=10.3,
            ytick distance=1,
            xlabel={$p_e$},
            ylabel={Average number of steps},
            grid=both,
            grid style={line width=.1pt, draw=gray!10},
            major grid style={line width=.2pt,draw=gray!50},
            every axis/.append style={
                x label style={
                    font=\fontsize{8}{8}\selectfont,
                    at={(axis description cs:0.5,-0.1)},
                    },
                y label style={
                    font=\fontsize{8}{8}\selectfont,
                    at={(axis description cs:-0.1,0.5)},
                    },
                x tick label style={
                    font=\fontsize{8}{8}\selectfont,
                    /pgf/number format/.cd,
                    fixed,
                    fixed zerofill,
                    precision=2,
                    /tikz/.cd
                    },
                y tick label style={
                    font=\fontsize{8}{8}\selectfont,
                    /pgf/number format/.cd,
                    fixed,
                    fixed zerofill,
                    precision=1,
                    /tikz/.cd
                    },
            }
        }
        \begin{axis}[mark options={solid}]
        \addplot[blue, solid, line width=0.9pt, mark=triangle*, mark options={fill=blue, scale=1.6}] 
                table [x=PE, y=BINARY_NOISY, col sep=comma] {Data/bsc_grid_world.csv};
        \addlegendentry{Joint learning and communication ($M=7$)}
        
        \addplot[color=gray, dashed, line width=1.2pt, mark=*, mark options={fill=gray, solid, scale=1.1}] 
        table [x=PE, y=HAMMING_NOISY (HC), col sep=comma] {Data/bsc_grid_world.csv};
        \addlegendentry{Separate learning and communication / HC}
        
        \addplot[color=orange, solid, line width=0.7pt, mark=*, mark options={fill=orange, scale=1}] 
        table [x=PE, y=HAMMING_NOISY (RC), col sep=comma] {Data/bsc_grid_world.csv};
        \addlegendentry{Separate learning and communication / RC}
        
        \addplot[color=cyan, dashed, line width=1.2pt, mark=square*, mark options={fill=cyan, solid, scale=1.1}] 
        table [x=PE, y=OPT_NOISY (HC), col sep=comma] {Data/bsc_grid_world.csv};
        \addlegendentry{Optimal actions with Hamming code / HC}
        
        \addplot[color=magenta, solid, line width=0.7pt, mark=square*, mark options={fill=magenta, scale=0.8}] 
        table [x=PE, y=OPT_NOISY (RC), col sep=comma] {Data/bsc_grid_world.csv};
        \addlegendentry{Optimal actions with Hamming code / RC}
        
        \addplot[color=darkgray, solid, thick, mark=x, mark options={fill=darkgray, scale=1}] 
        table [x=PE, y=LB, col sep=comma] {Data/bsc_grid_world.csv};
        \addlegendentry{Optimal actions without noise}
        \end{axis}
        \end{tikzpicture}
        }
  \caption{Comparison of agents jointly trained to collaborate and communicate over a BSC to separate learning and communications with a (7,4) Hamming code.}
  \label{fig:bsc_grid_world} 
\vspace{-0.8cm}
\end{figure}

\begin{figure} 
    \centering
  \subfloat[$\delta=0$ \label{subfig:awgn_grid_world}]{%
    \begin{tikzpicture}
        \pgfplotsset{
            legend style={
                font=\fontsize{5.8}{5.8}\selectfont,
                at={(0.99,.99)},
                anchor=north east,
            },
            height=0.4\linewidth,
            width=0.5\linewidth,
            xmin=0,
            xmax=23,
            xtick distance=3,
            ymin=2.3,
            ymax=4.3,
            ytick distance=0.5,
            xlabel={SNR (dB)},
            ylabel={Average number of steps},
            grid=both,
            grid style={line width=.1pt, draw=gray!10},
            major grid style={line width=.2pt,draw=gray!50},
            every axis/.append style={
                x label style={
                    font=\fontsize{8}{8}\selectfont,
                    at={(axis description cs:0.5,-0.1)},
                    },
                y label style={
                    font=\fontsize{8}{8}\selectfont,
                    at={(axis description cs:-0.1,0.5)},
                    },
                x tick label style={
                    font=\fontsize{8}{8}\selectfont,
                    /pgf/number format/.cd,
                    fixed,
                    fixed zerofill,
                    precision=2,
                    /tikz/.cd
                    },
                y tick label style={
                    font=\fontsize{8}{8}\selectfont,
                    /pgf/number format/.cd,
                    fixed,
                    fixed zerofill,
                    precision=1,
                    /tikz/.cd
                    },
            }
        }
        \begin{axis}
        \addplot[color=blue, solid, line width=0.9pt, mark=triangle*, mark options={fill=blue, scale=1.6}] 
                table [x=SNR, y=BINARY, col sep=comma] {Data/awgn_grid_world.csv};
        \addlegendentry{Joint learning and communication (BPSK, $M=7$)}
        
        \addplot[color=blue, dashed, line width=1.2pt, mark=triangle*, mark options={fill=blue, solid, scale=1.6}] 
                table [x=SNR, y=REAL, col sep=comma] {Data/awgn_grid_world.csv};
        \addlegendentry{Joint learning and communication (Real, $M=7$)}
        
        \addplot[color=gray, dashed, line width=1.2pt, mark=*, mark options={fill=gray, solid, scale=1.1}] 
        table [x=SNR, y=HAMMING (HC), col sep=comma] {Data/awgn_grid_world.csv};
        \addlegendentry{Separate learning and communication / HC}
        
        \addplot[color=orange, solid, line width=0.7pt, mark=*, mark options={fill=orange, scale=1}] 
        table [x=SNR, y=HAMMING (RC), col sep=comma] {Data/awgn_grid_world.csv};
        \addlegendentry{Separate learning and communication / RC}
        
        \addplot[color=cyan, dashed, line width=1.2pt, mark=square*, mark options={fill=cyan, solid, scale=1.1}] 
        table [x=SNR, y=OPT (HC), col sep=comma] {Data/awgn_grid_world.csv};
        \addlegendentry{Optimal with Hamming code / HC}
        
        \addplot[color=magenta, solid, line width=0.7pt, mark=square*, mark options={fill=magenta, scale=1}] 
        table [x=SNR, y=OPT (RC), col sep=comma] {Data/awgn_grid_world.csv};
        \addlegendentry{Optimal with Hamming code / RC}
        
        \addplot[color=darkgray, solid, thick, mark=x, mark options={fill=darkgray, scale=1}] 
        table [x=SNR, y=LB, col sep=comma] {Data/awgn_grid_world.csv};
        \addlegendentry{Optimal actions without noise}
        \end{axis}
        \end{tikzpicture}
    }
  \subfloat[$\delta=0.05$ \label{subfig:awgn_grid_world_noisy}]{%
    \begin{tikzpicture}
        \pgfplotsset{
            legend style={
                font=\fontsize{5.8}{5.8}\selectfont,
                at={(0.99,.99)},
                anchor=north east,
            },
            height=0.4\linewidth,
            width=0.5\linewidth,
            xmin=0,
            xmax=23,
            xtick distance=3,
            ymin=2.,
            ymax=5.5,
            ytick distance=0.5,
            xlabel={SNR (dB)},
            ylabel={Average number of steps},
            grid=both,
            grid style={line width=.1pt, draw=gray!10},
            major grid style={line width=.2pt,draw=gray!50},
            every axis/.append style={
                x label style={
                    font=\fontsize{8}{8}\selectfont,
                    at={(axis description cs:0.5,-0.1)},
                    },
                y label style={
                    font=\fontsize{8}{8}\selectfont,
                    at={(axis description cs:-0.1,0.5)},
                    },
                x tick label style={
                    font=\fontsize{8}{8}\selectfont,
                    /pgf/number format/.cd,
                    fixed,
                    fixed zerofill,
                    precision=2,
                    /tikz/.cd
                    },
                y tick label style={
                    font=\fontsize{8}{8}\selectfont,
                    /pgf/number format/.cd,
                    fixed,
                    fixed zerofill,
                    precision=1,
                    /tikz/.cd
                    },
            }
        }
        \begin{axis}[mark options={solid}]
        \addplot[color=blue, solid, line width=0.9pt, mark=triangle*, mark options={fill=blue, scale=1.6}] 
                table [x=SNR, y=BINARY, col sep=comma] {Data/awgn_grid_world_noisy.csv};
        \addlegendentry{Joint learning and communication (BPSK, $M=7$)}
        
        \addplot[color=blue, dashed, line width=1.2pt, mark=triangle*, mark options={fill=blue, solid, scale=1.6}] 
                table [x=SNR, y=REAL, col sep=comma] {Data/awgn_grid_world_noisy.csv};
        \addlegendentry{Joint learning and communication (Real, $M=7$)}
        
        \addplot[color=gray, dashed, line width=1.2pt, mark=*, mark options={fill=gray, solid, scale=1.1}] 
        table [x=SNR, y=HAMMING (HC), col sep=comma] {Data/awgn_grid_world_noisy.csv};
        \addlegendentry{Separate learning and communication / HC}
        
        \addplot[color=orange, solid, line width=0.7pt, mark=*, mark options={fill=orange, scale=1}] 
        table [x=SNR, y=HAMMING (RC), col sep=comma] {Data/awgn_grid_world_noisy.csv};
        \addlegendentry{Separate learning and communication / RC}
        
        \addplot[color=cyan, dashed, line width=1.2pt, mark=square*, mark options={fill=cyan, solid, scale=1.1}] 
        table [x=SNR, y=OPT (HC), col sep=comma] {Data/awgn_grid_world_noisy.csv};
        \addlegendentry{Optimal actions with Hamming code / HC}
        
        \addplot[color=magenta, solid, line width=0.7pt, mark=square*, mark options={fill=magenta, scale=1}] 
        table [x=SNR, y=OPT (RC), col sep=comma] {Data/awgn_grid_world_noisy.csv};
        \addlegendentry{Optimal actions with Hamming code / RC}
        
        \addplot[color=darkgray, solid, thick, mark=x, mark options={fill=darkgray, scale=1}] 
        table [x=SNR, y=LB, col sep=comma] {Data/awgn_grid_world_noisy.csv};
        \addlegendentry{Optimal actions without noise}
        \end{axis}
        \end{tikzpicture}
        }
  \caption{Comparison of the agents jointly trained to collaborate and communicate over an AWGN channel to separate learning and communications with a (7,4) Hamming code.}
  \label{fig:awgn_grid_world} 
\vspace{-1cm}
\end{figure}
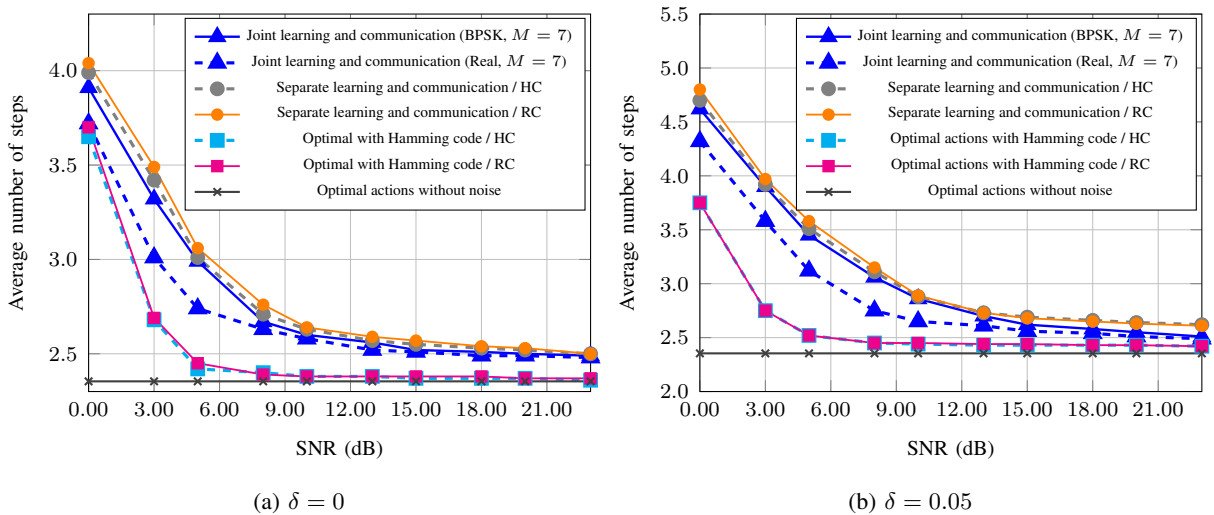

\begin{figure} 
    \centering
  \subfloat[Separate learning and communication (HC). \label{subfig:hamming_bsc_vis}]{%
    \includegraphics[height=0.2\linewidth]{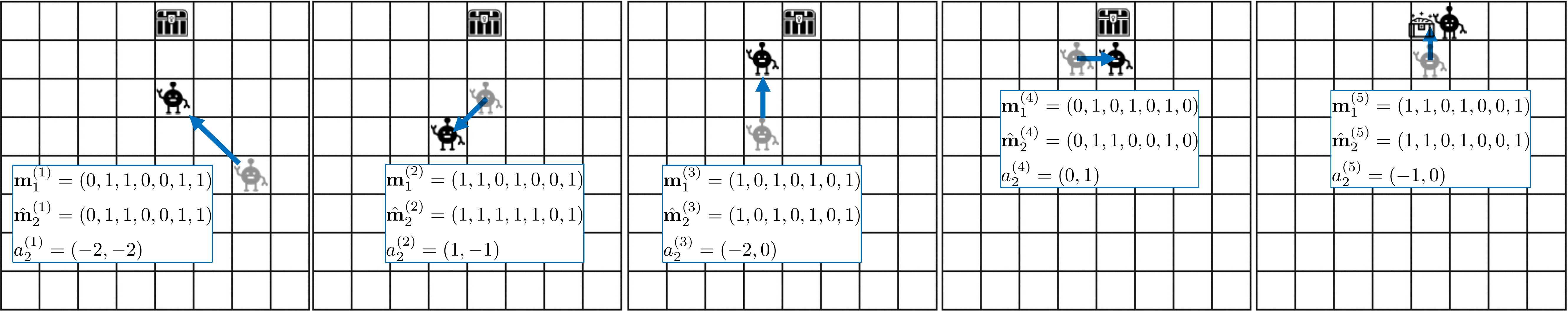}
    }\\
  \subfloat[Joint learning and communication. \label{subfig:learning_bsc_vis}]{%
    \includegraphics[height=0.2\linewidth]{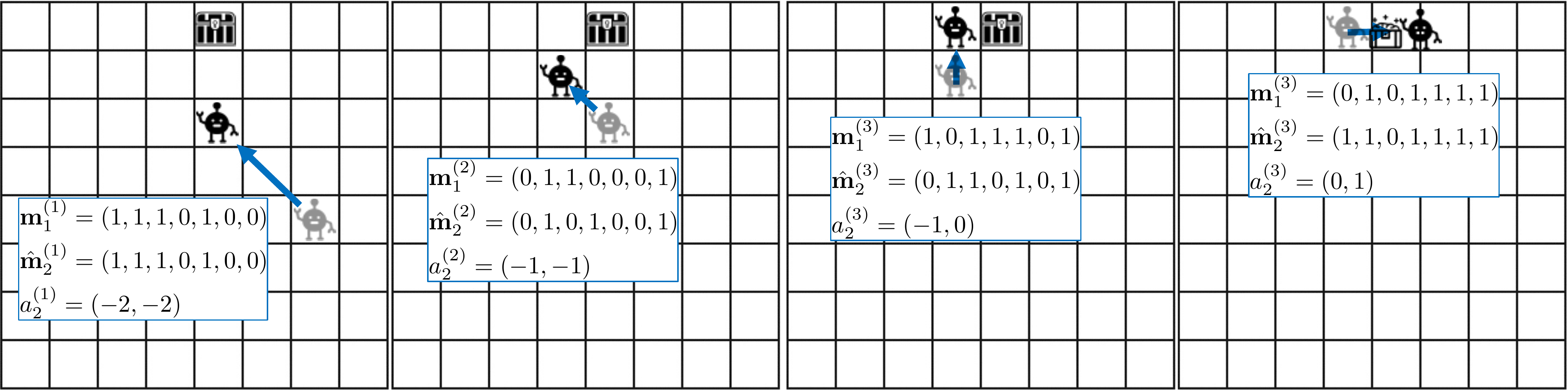}
        }
  \caption{Example visualization of the codewords used by the guide, and the path taken by the scout for $M=7$ uses of a BSC with $p_e=0.2$ and $\delta=0$. The origin is at the top left corner.}
  \label{fig:bsc_vis} 
\vspace{-0.8cm}
\end{figure}

For the grid world problem, presented in Section \ref{subsec:eg_prob_guide_scout}, the scout and treasure are uniformly randomly placed on any distinct locations upon initialization (i.e., $p_g\ne p_s^{(0)}$).
These locations are one-hot encoded to form a $2L^2$ vector that is the observation of the guide $o_1^{(t)}$. 
We fix the channel bandwidth to $M=\{7,10\}$ and compare our solutions to a scheme that separates the channel coding from the underlying MDP.
That is, we first train a RL agent that solves the grid world problem without communication constraints. 
We then introduce a noisy communication channel and encode the action chosen by the RL agent using a (7,4) Hamming code before transmission across the channel.
The received message is then decoded and the resultant action is taken.
We note that the (7,4) Hamming code is a perfect code that encodes four data bits into seven channel bits by adding three parity bits; thus, it can correct single bit errors.
The association between the 16 possible actions and codewords of 4 bits can be done by random permutation, which we refer to as random codewords (RC), or hand-crafted (HC) association by assigning adjacent codewords to similar actions, as shown in Fig. \ref{fig:grid_world}.
By associating adjacent codewords to similar actions, the scout will take a similar action to the one intended even if there is a decoding error, assuming the number of bit errors is not too high.
Lastly, we compute the optimal solution, where the steps taken forms the shortest path to the treasure, and use a Hamming (7,4) channel code to transmit those actions.
This is referred to as ``Optimal actions with Hamming Code" and acts as a lower bound for the separation-based results.

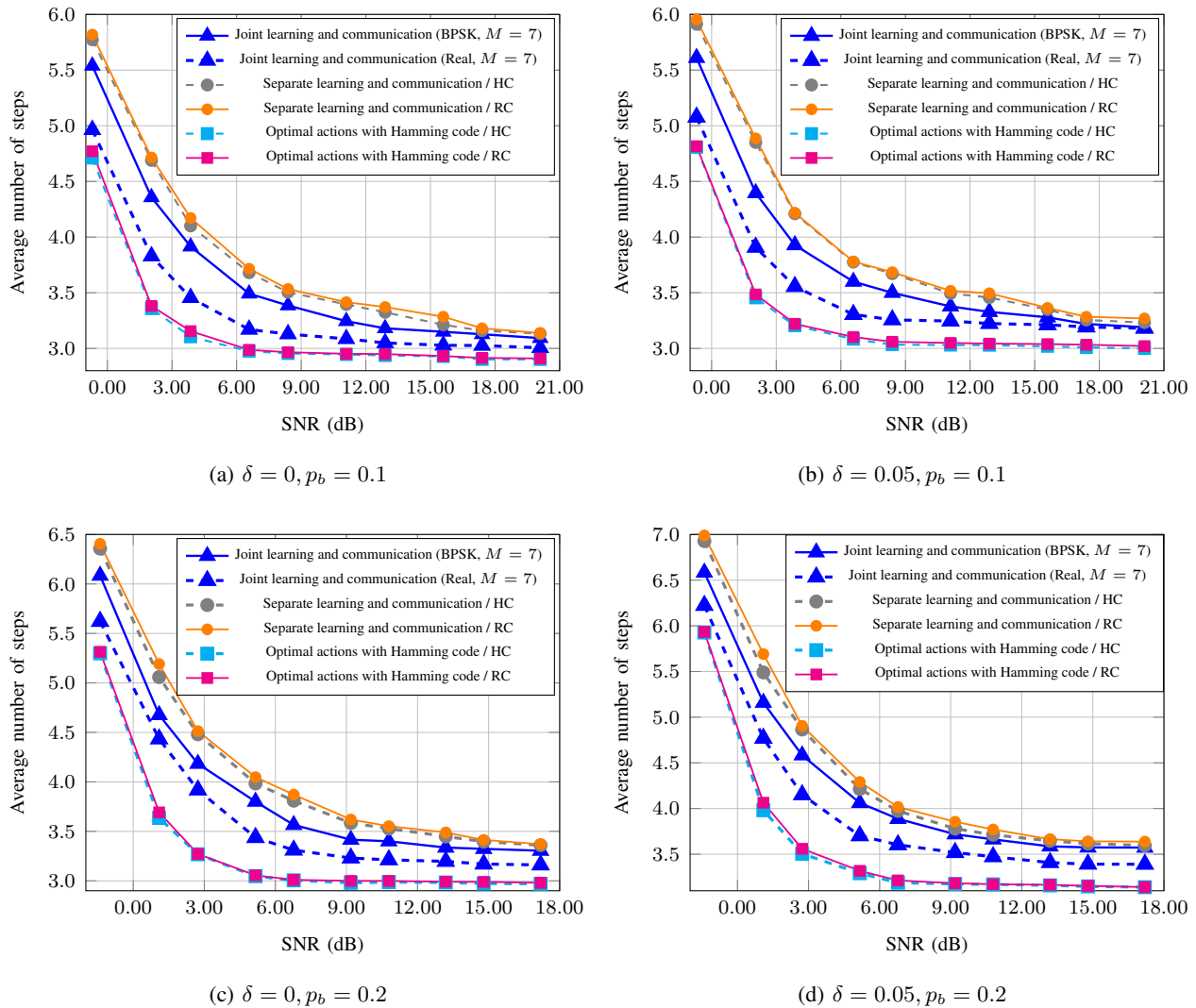
\begin{figure} 
    \centering
  \subfloat[$\delta=0,p_b=0.1$ \label{subfig:bn01_grid_world}]{%
    \begin{tikzpicture}
        \pgfplotsset{
            legend style={
                font=\fontsize{5.8}{5.8}\selectfont,
                at={(0.99,.99)},
                anchor=north east,
            },
            height=0.4\linewidth,
            width=0.5\linewidth,
            xmin=-1,
            xmax=21,
            xtick distance=3,
            ymin=2.8,
            ymax=6,
            ytick distance=0.5,
            xlabel={SNR (dB)},
            ylabel={Average number of steps},
            grid=both,
            grid style={line width=.1pt, draw=gray!10},
            major grid style={line width=.2pt,draw=gray!50},
            every axis/.append style={
                x label style={
                    font=\fontsize{8}{8}\selectfont,
                    at={(axis description cs:0.5,-0.1)},
                    },
                y label style={
                    font=\fontsize{8}{8}\selectfont,
                    at={(axis description cs:-0.1,0.5)},
                    },
                x tick label style={
                    font=\fontsize{8}{8}\selectfont,
                    /pgf/number format/.cd,
                    fixed,
                    fixed zerofill,
                    precision=2,
                    /tikz/.cd
                    },
                y tick label style={
                    font=\fontsize{8}{8}\selectfont,
                    /pgf/number format/.cd,
                    fixed,
                    fixed zerofill,
                    precision=1,
                    /tikz/.cd
                    },
            }
        }
        \begin{axis}[mark options={solid}]
        \addplot[blue, solid, line width=0.9pt, mark=triangle*, mark options={fill=blue, scale=1.5}] 
                table [x=SNR, y=BINARY, col sep=comma] {Data/bn_grid_world_p01.csv};
        \addlegendentry{Joint learning and communication (BPSK, $M=7$)}
        
        \addplot[blue, dashed, line width=1.2pt, mark=triangle*, mark options={fill=blue, solid, scale=1.5}] 
                table [x=SNR, y=REAL, col sep=comma] {Data/bn_grid_world_p01.csv};
        \addlegendentry{Joint learning and communication (Real, $M=7$)}
        
        \addplot[color=gray, dashed, line width=1.2pt, thick, mark=*, mark options={fill=gray, solid, scale=1.1}] 
        table [x=SNR, y=HAMMING (HC), col sep=comma] {Data/bn_grid_world_p01.csv};
        \addlegendentry{Separate learning and communication / HC}
        
        \addplot[color=orange, solid, line width=0.7pt, mark=*, mark options={fill=orange, scale=1}] 
        table [x=SNR, y=HAMMING (RC), col sep=comma] {Data/bn_grid_world_p01.csv};
        \addlegendentry{Separate learning and communication / RC}
        
        \addplot[color=cyan, dashed, line width=1.2pt, thick, mark=square*, mark options={fill=cyan, solid, scale=1.1}] 
        table [x=SNR, y=OPT (HC), col sep=comma] {Data/bn_grid_world_p01.csv};
        \addlegendentry{Optimal actions with Hamming code / HC}
        
        \addplot[color=magenta, solid, line width=0.7pt, mark=square*, mark options={fill=magenta, scale=1}] 
        table [x=SNR, y=OPT (RC), col sep=comma] {Data/bn_grid_world_p01.csv};
        \addlegendentry{Optimal actions with Hamming code / RC}
        
        \end{axis}
        \end{tikzpicture}
       }
  \subfloat[$\delta=0.05,p_b=0.1$ \label{subfig:bn01_grid_world_noisy}]{%
    \begin{tikzpicture}
        \pgfplotsset{
            legend style={
                font=\fontsize{5.8}{5.8}\selectfont,
                at={(0.99,.99)},
                anchor=north east,
            },
            height=0.4\linewidth,
            width=0.5\linewidth,
            xmin=-1,
            xmax=21,
            xtick distance=3,
            ymin=2.8,
            ymax=6,
            ytick distance=0.5,
            xlabel={SNR (dB)},
            ylabel={Average number of steps},
            grid=both,
            grid style={line width=.1pt, draw=gray!10},
            major grid style={line width=.2pt,draw=gray!50},
            every axis/.append style={
                x label style={
                    font=\fontsize{8}{8}\selectfont,
                    at={(axis description cs:0.5,-0.1)},
                    },
                y label style={
                    font=\fontsize{8}{8}\selectfont,
                    at={(axis description cs:-0.1,0.5)},
                    },
                x tick label style={
                    font=\fontsize{8}{8}\selectfont,
                    /pgf/number format/.cd,
                    fixed,
                    fixed zerofill,
                    precision=2,
                    /tikz/.cd
                    },
                y tick label style={
                    font=\fontsize{8}{8}\selectfont,
                    /pgf/number format/.cd,
                    fixed,
                    fixed zerofill,
                    precision=1,
                    /tikz/.cd
                    },
            }
        }
        \begin{axis}[mark options={solid}]
        \addplot[blue, solid, line width=0.9pt, mark=triangle*, mark options={fill=blue, scale=1.6}] 
                table [x=SNR, y=BINARY, col sep=comma] {Data/bn_grid_world_noisy_p01.csv};
        \addlegendentry{Joint learning and communication (BPSK, $M=7$)}
        
        \addplot[blue, dashed, line width=1.2pt, mark=triangle*, mark options={fill=blue, solid, scale=1.6}] 
                table [x=SNR, y=REAL, col sep=comma] {Data/bn_grid_world_noisy_p01.csv};
        \addlegendentry{Joint learning and communication (Real, $M=7$)}
        
        \addplot[color=gray, dashed, line width=1.2pt, thick, mark=*, mark options={fill=gray, solid, scale=1.1}] 
        table [x=SNR, y=HAMMING (HC), col sep=comma] {Data/bn_grid_world_noisy_p01.csv};
        \addlegendentry{Separate learning and communication / HC}
        
        \addplot[color=orange, solid, line width=0.7pt, mark=*, mark options={fill=orange, scale=1}] 
        table [x=SNR, y=HAMMING (RC), col sep=comma] {Data/bn_grid_world_noisy_p01.csv};
        \addlegendentry{Separate learning and communication / RC}
        
        \addplot[color=cyan, dashed, line width=1.2pt, thick, mark=square*, mark options={fill=cyan, solid, scale=1.1}] 
        table [x=SNR, y=OPT (HC), col sep=comma] {Data/bn_grid_world_noisy_p01.csv};
        \addlegendentry{Optimal actions with Hamming code / HC}
        
        \addplot[color=magenta, solid, line width=0.7pt, mark=square*, mark options={fill=magenta, scale=1}] 
        table [x=SNR, y=OPT (RC), col sep=comma] {Data/bn_grid_world_noisy_p01.csv};
        \addlegendentry{Optimal actions with Hamming code / RC}
        
        \end{axis}
        \end{tikzpicture}
        }
        \\
  \subfloat[$\delta=0,p_b=0.2$ \label{subfig:bn02_grid_world}]{%
    \begin{tikzpicture}
        \pgfplotsset{
            legend style={
                font=\fontsize{5.8}{5.8}\selectfont,
                at={(0.99,.99)},
                anchor=north east,
            },
            height=0.4\linewidth,
            width=0.5\linewidth,
            xmin=-2,
            xmax=18,
            xtick distance=3,
            ymin=2.9,
            ymax=6.5,
            ytick distance=0.5,
            xlabel={SNR (dB)},
            ylabel={Average number of steps},
            grid=both,
            grid style={line width=.1pt, draw=gray!10},
            major grid style={line width=.2pt,draw=gray!50},
            every axis/.append style={
                x label style={
                    font=\fontsize{8}{8}\selectfont,
                    at={(axis description cs:0.5,-0.1)},
                    },
                y label style={
                    font=\fontsize{8}{8}\selectfont,
                    at={(axis description cs:-0.1,0.5)},
                    },
                x tick label style={
                    font=\fontsize{8}{8}\selectfont,
                    /pgf/number format/.cd,
                    fixed,
                    fixed zerofill,
                    precision=2,
                    /tikz/.cd
                    },
                y tick label style={
                    font=\fontsize{8}{8}\selectfont,
                    /pgf/number format/.cd,
                    fixed,
                    fixed zerofill,
                    precision=1,
                    /tikz/.cd
                    },
            }
        }
        \begin{axis}[mark options={solid}]
        \addplot[blue, solid, line width=0.9pt, mark=triangle*, mark options={fill=blue, scale=1.6}] 
                table [x=SNR, y=BINARY, col sep=comma] {Data/bn_grid_world_p02.csv};
        \addlegendentry{Joint learning and communication (BPSK, $M=7$)}
        
        \addplot[blue, dashed, line width=1.2pt, mark=triangle*, mark options={fill=blue, solid, scale=1.6}] 
                table [x=SNR, y=REAL, col sep=comma] {Data/bn_grid_world_p02.csv};
        \addlegendentry{Joint learning and communication (Real, $M=7$)}
        
        \addplot[color=gray, dashed, line width=1.2pt, mark=*, mark options={fill=gray, solid, scale=1.1}] 
        table [x=SNR, y=HAMMING (HC), col sep=comma] {Data/bn_grid_world_p02.csv};
        \addlegendentry{Separate learning and communication / HC}
        
        \addplot[color=orange, solid, line width=0.7pt, mark=*, mark options={fill=orange, scale=1}] 
        table [x=SNR, y=HAMMING (RC), col sep=comma] {Data/bn_grid_world_p02.csv};
        \addlegendentry{Separate learning and communication / RC}
        
        \addplot[color=cyan, dashed, line width=1.2pt, mark=square*, mark options={fill=cyan, solid, scale=1.1}] 
        table [x=SNR, y=OPT (HC), col sep=comma] {Data/bn_grid_world_p02.csv};
        \addlegendentry{Optimal actions with Hamming code / HC}
        
        \addplot[color=magenta, solid, line width=0.7pt, mark=square*, mark options={fill=magenta, scale=1}] 
        table [x=SNR, y=OPT (RC), col sep=comma] {Data/bn_grid_world_p02.csv};
        \addlegendentry{Optimal actions with Hamming code / RC}
        
        \end{axis}
        \end{tikzpicture}
       }
  \subfloat[$\delta=0.05,p_b=0.2$ \label{subfig:bn02_grid_world_noisy}]{%
    \begin{tikzpicture}
        \pgfplotsset{
            legend style={
                font=\fontsize{5.8}{5.8}\selectfont,
                at={(1.0,1.)},
                anchor=north east,
            },
            height=0.4\linewidth,
            width=0.5\linewidth,
            xmin=-2,
            xmax=18,
            xtick distance=3,
            ymin=3.1,
            ymax=7,
            ytick distance=0.5,
            xlabel={SNR (dB)},
            ylabel={Average number of steps},
            grid=both,
            grid style={line width=.1pt, draw=gray!10},
            major grid style={line width=.2pt,draw=gray!50},
            every axis/.append style={
                x label style={
                    font=\fontsize{8}{8}\selectfont,
                    at={(axis description cs:0.5,-0.1)},
                    },
                y label style={
                    font=\fontsize{8}{8}\selectfont,
                    at={(axis description cs:-0.1,0.5)},
                    },
                x tick label style={
                    font=\fontsize{8}{8}\selectfont,
                    /pgf/number format/.cd,
                    fixed,
                    fixed zerofill,
                    precision=2,
                    /tikz/.cd
                    },
                y tick label style={
                    font=\fontsize{8}{8}\selectfont,
                    /pgf/number format/.cd,
                    fixed,
                    fixed zerofill,
                    precision=1,
                    /tikz/.cd
                    },
            }
        }
        \begin{axis}[mark options={solid}]
        \addplot[blue, solid, line width=0.9pt, mark=triangle*, mark options={fill=blue, scale=1.6}] 
                table [x=SNR, y=BINARY, col sep=comma] {Data/bn_grid_world_noisy_p02.csv};
        \addlegendentry{Joint learning and communication (BPSK, $M=7$)}
        
        \addplot[blue, dashed, line width=1.2pt, mark=triangle*, mark options={fill=blue, solid, scale=1.6}] 
                table [x=SNR, y=REAL, col sep=comma] {Data/bn_grid_world_noisy_p02.csv};
        \addlegendentry{Joint learning and communication (Real, $M=7$)}
        
        \addplot[color=gray, dashed, line width=1.2pt, mark=*, mark options={fill=gray, solid, scale=1.1}] 
        table [x=SNR, y=HAMMING (HC), col sep=comma] {Data/bn_grid_world_noisy_p02.csv};
        \addlegendentry{Separate learning and communication / HC}
        
        \addplot[color=orange, solid, line width=0.7pt, mark=*, mark options={fill=orange, scale=1}] 
        table [x=SNR, y=HAMMING (RC), col sep=comma] {Data/bn_grid_world_noisy_p02.csv};
        \addlegendentry{Separate learning and communication / RC}
        
        \addplot[color=cyan, dashed, line width=1.2pt, mark=square*, mark options={fill=cyan, solid, scale=1.1}] 
        table [x=SNR, y=OPT (HC), col sep=comma] {Data/bn_grid_world_noisy_p02.csv};
        \addlegendentry{Optimal actions with Hamming code / HC}
        
        \addplot[color=magenta, solid, line width=0.7pt, mark=square*, mark options={fill=magenta, scale=1}] 
        table [x=SNR, y=OPT (RC), col sep=comma] {Data/bn_grid_world_noisy_p02.csv};
        \addlegendentry{Optimal actions with Hamming code / RC}
        
        \end{axis}
        \end{tikzpicture}
        }
  \caption{Comparison of the agents jointly trained to collaborate and communicate over an BN channel to separate learning and communications with a (7,4) Hamming code.}
  \label{fig:bn_grid_world} 
\vspace{-1.0cm}
\end{figure}

For the joint channel coding-modulation problem, we again compare the DDPG and actor-critic results with a (7,4) Hamming code using BPSK modulation.
The source bit sequence is uniformly randomly chosen from the set $\{0,1\}^M$ and one-hot encoded to form the input state $o_1^{(1)}$ of the transmitter.
We also compare with the algorithm derived in \cite{aoudia_model-free_2019}, which uses supervised learning for the receiver and the REINFORCE policy gradient to estimate the gradient of the transmitter.



We first present the results for the guided robot problem. 
Fig. \ref{fig:bsc_grid_world} shows the number of steps, averaged over 10K episodes, needed by the scout to reach the treasure for the BSC case with $\delta=\{0,0.05\}$. 
The ``optimal actions without noise" refers to the minimum number of steps required to reach the treasure assuming a perfect communication channel and acts as the lower bound for all the experiments.
It is clear that jointly learning to communicate and collaborate over a noisy channel outperforms the separation-based results with both RC and HC.
In Fig. \ref{fig:bsc_vis}, we provide an illustration of the actions taken by the agent after some errors over the communication channel with the separate learning and communication scheme (HC) and with the proposed joint learning and communication approach. It can be seen that at step 2 the proposed scheme takes a similar action $(-1,-1)$ to the optimal one $(-2,0)$ despite experiencing 2 bit errors, and in step 3 despite experiencing 3 bit errors (Fig. \ref{subfig:learning_bsc_vis}). On the other hand, in the separate learning and communication scheme with a (7,4) Hamming code and HC association of actions, the scout decodes a very different action from the optimal one in step 2 which results in an additional step being taken. However, it was able to take a similar action to the optimal one in step 4 despite experiencing 2 bit errors. This shows that although hand crafting codeword assignments can lead to some performance benefits in the separate learning and communication scheme, which was also suggested by Fig. \ref{fig:bsc_grid_world}, joint learning and communication leads to more robust codeword assignments that give much more consistent results. Indeed, we have also observed that, unlike the separation based scheme, where each message corresponds to a single action, or equivalently, there are 8 different channel output vectors for which the same action is taken, the codeword to action mapping at the scout can be highly asymmetric for the learned scheme. 
Moreover, neither the joint learning and communication results nor the separation-based results achieve the performance of the optimal solution with Hamming code. The gap between the optimal solution with Hamming code and the results obtained by the guide/scout formulation is due to the DQN architectures' limited capability to learn the optimal solution and the challenge of learning under noisy environments.
Comparing Fig. \ref{subfig:bsc_grid_world} and \ref{subfig:bsc_grid_world_noisy}, the performance degradation due to the separation-based results is slightly greater than those from the joint framework. This is because the joint learning and communication approach is better at adjusting its policy and communication strategy to mitigate the effect of the channel noise than employing a standard channel code.

\begin{figure}
\begin{minipage}{.48\textwidth}
  \centering
    \begin{tikzpicture}
        \pgfplotsset{
            legend style={
                font=\fontsize{6}{6}\selectfont,
                at={(0.92,.98)},
                anchor=north east,
            },
            height=0.8\linewidth,
            width=\linewidth,
            xmin=0,
            xmax=100000,
            ymin=0,
            ymax=150,
            xlabel={Episode},
            ylabel={Number of steps},
            grid=both,
            grid style={line width=.1pt, draw=gray!10},
            major grid style={line width=.2pt,draw=gray!50},
            every axis/.append style={
                x label style={
                    font=\fontsize{8}{8}\selectfont,
                    at={(axis description cs:0.5,-0.1)},
                    },
                y label style={
                    font=\fontsize{8}{8}\selectfont,
                    at={(axis description cs:-0.12,0.5)},
                    },
                x tick label style={
                    font=\fontsize{8}{8}\selectfont,
                    /pgf/number format/.cd,
                    fixed,
                    fixed zerofill,
                    precision=2,
                    /tikz/.cd
                    },
                y tick label style={
                    font=\fontsize{8}{8}\selectfont,
                    /pgf/number format/.cd,
                    fixed,
                    fixed zerofill,
                    precision=1,
                    /tikz/.cd
                    },
            },
        }
        \begin{axis}
        \addplot[no markers, blue] table [x=Step, y=BPSK BSC, col sep=comma] {Data/conv_mdp.csv};
        \addlegendentry{BPSK BSC ($P_e=0.05$)}
        
        \addplot[no markers, red] table [x=Step, y=REAL AWGN, col sep=comma] {Data/conv_mdp.csv};
        \addlegendentry{Real AWGN ($10$ dB)}
        
        \addplot[no markers, green] table [x=Step, y=BPSK AWGN, col sep=comma] {Data/conv_mdp.csv};
        \addlegendentry{BPSK AWGN ($10$ dB)}
        \end{axis}
    \end{tikzpicture}
  \caption{Convergence of each channel scenario for the grid world problem without noise ($M=7,~\delta=0$).}
  \label{fig:mdp_convergence}
  \vspace{0.2cm}
\end{minipage}%
\hfill
\begin{minipage}{.48\textwidth}
  \centering
    \begin{tikzpicture}
        \pgfplotsset{
            legend style={
                font=\fontsize{6}{6}\selectfont,
                at={(0.99,.99)},
                anchor=north east,
            },
            height=0.8\linewidth,
            width=\linewidth,
            xmin=0,
            xmax=23,
            ymin=2.3,
            ymax=4.1,
            xlabel={SNR (dB)},
            ylabel={Average number of steps},
            grid=both,
            grid style={line width=.1pt, draw=gray!10},
            major grid style={line width=.2pt,draw=gray!50},
            every axis/.append style={
                x label style={
                    font=\fontsize{8}{8}\selectfont,
                    at={(axis description cs:0.5,-0.1)},
                    },
                y label style={
                    font=\fontsize{8}{8}\selectfont,
                    at={(axis description cs:-0.1,0.5)},
                    },
                x tick label style={
                    font=\fontsize{8}{8}\selectfont,
                    /pgf/number format/.cd,
                    fixed,
                    fixed zerofill,
                    precision=2,
                    /tikz/.cd
                    },
                y tick label style={
                    font=\fontsize{8}{8}\selectfont,
                    /pgf/number format/.cd,
                    fixed,
                    fixed zerofill,
                    precision=1,
                    /tikz/.cd
                    },
            }
        }
        \begin{axis}
        \addplot[blue, solid, line width=0.9pt, mark=triangle*, mark options={fill=blue, scale=1.5}] 
                table [x=SNR, y=BINARY, col sep=comma] {Data/awgn_grid_world.csv};
        \addlegendentry{Joint learning and communication (BPSK, $M=7$)}
        
        \addplot[blue, dashed, line width=1.2pt, mark=triangle*, mark options={fill=blue, solid, scale=1.5}] 
                table [x=SNR, y=REAL, col sep=comma] {Data/awgn_grid_world.csv};
        \addlegendentry{Joint learning and communication (Real, $M=7$)}
        
        \addplot[orange, solid, line width=0.9pt, mark=triangle*, mark options={fill=orange, scale=1.5}] 
                table [x=SNR, y=BINARY, col sep=comma] {Data/awgn_grid_world_m10.csv};
        \addlegendentry{Joint learning and communication (BPSK, $M=10$)}
        
        \addplot[orange, dashed, line width=1.2pt, mark=triangle*, mark options={fill=orange, solid, scale=1.5}] 
                table [x=SNR, y=REAL, col sep=comma] {Data/awgn_grid_world_m10.csv};
        \addlegendentry{Joint learning and communication (Real, $M=10$)}
        
        \end{axis}
        \end{tikzpicture}
  \caption{Impact of the channel bandwidth $M=\{7,10\}$ on the performance for an AWGN channel ($\delta=0$).}
  \label{fig:bw_affec}
\end{minipage}
\vspace{-1cm}
\end{figure}

Similarly, in the AWGN case in Fig. \ref{fig:awgn_grid_world}, the results from joint learning and communication clearly outperforms those obtained via separate learning and communication.
Here, the ``Real" results refer to the guide agent with $\mathcal{A}_1=\mathbb{R}^M$, while the ``BPSK" results refer to the guide agent with $\mathcal{A}_1=\{-1,+1\}^M$.
The ``Real" results here clearly outperform all other schemes considered. The relaxation of the channel constellation to all real values within a power constraint allows the guide to convey more information than a binary constellation can achieve. We also observe that the gain from this relaxation is higher at lower SNR values for both $\delta$ values. This is in contrast to the gap between the channel capacities achieved with Gaussian and binary inputs in an AWGN channel, which is negligible at low SNR values and increases with SNR. This shows that channel capacity is not the right metric for this problem, and even when two channels are similar in terms of capacity, they can give very different performances in terms of the discounted sum reward when used in the MARL context.

\begin{figure}
\begin{minipage}{.48\textwidth}
  \centering
    \begin{tikzpicture}
        \pgfplotsset{
            legend style={
                font=\fontsize{6}{6}\selectfont,
                at={(0.4,.35)},
                anchor=north east,
            },
            height=0.8\linewidth,
            width=\linewidth,
            xmin=0,
            xmax=5,
            ymin=0.001,
            ymax=0.4,
            xlabel={SNR (dB)},
            ylabel={BLER},
            grid=both,
            grid style={line width=.1pt, draw=gray!10},
            major grid style={line width=.2pt,draw=gray!50},
            every axis/.append style={
                x label style={
                    font=\fontsize{8}{8}\selectfont,
                    at={(axis description cs:0.5,-0.1)},
                    },
                y label style={
                    font=\fontsize{8}{8}\selectfont,
                    at={(axis description cs:-0.1,0.5)},
                    },
                x tick label style={
                    font=\fontsize{8}{8}\selectfont,
                    /pgf/number format/.cd,
                    fixed,
                    fixed zerofill,
                    precision=2,
                    /tikz/.cd
                    },
                y tick label style={
                    font=\fontsize{8}{8}\selectfont,
                    /pgf/number format/.cd,
                    fixed,
                    fixed zerofill,
                    precision=1,
                    /tikz/.cd
                    },
            }
        }
        \begin{axis}[
            ymode=log,
            log ticks with fixed point,
        ]
        
        \addplot[blue, solid, line width=0.9pt, mark=triangle*, mark options={fill=blue, scale=1.5}] 
                table [x=SNR, y=HAMMING, col sep=comma] {Data/ch_coding_awgn.csv};
        \addlegendentry{HAMMING}
        
        \addplot[gray, dashed, line width=1.2pt, mark=square*, mark options={fill=gray, solid, scale=1.}] 
                table [x=SNR, y=DDPG, col sep=comma] {Data/ch_coding_awgn.csv};
        \addlegendentry{DDPG}
        
        \addplot[cyan, solid, line width=0.9pt, mark=*, mark options={fill=cyan, scale=1.}] 
                table [x=SNR, y=REINFORCE, col sep=comma] {Data/ch_coding_awgn.csv};
        \addlegendentry{REINFORCE}
        
        \addplot[magenta, dashed, line width=1.2pt, mark=x, mark options={fill=magenta, scale=1.5, solid}] 
                table [x=SNR, y=A2C, col sep=comma] {Data/ch_coding_awgn.csv};
        \addlegendentry{Actor-Critic}
        
        \end{axis}
        \end{tikzpicture}
  \caption{BLER performance of different modulation and coding schemes over AWGN channel.}
  \label{fig:ch_coding_bler}
  \vspace{-0.2cm}
\end{minipage}
\hfill
\begin{minipage}{.48\textwidth}
  \centering
    \begin{tikzpicture}
        \pgfplotsset{
            legend style={
                font=\fontsize{6}{6}\selectfont,
                at={(0.95,.98)},
                anchor=north east,
            },
            height=0.8\linewidth,
            width=\linewidth,
            xmin=0,
            xmax=10000,
            ymin=0.007,
            ymax=0.1,
            xlabel={Episode},
            ylabel={BLER},
            grid=both,
            grid style={line width=.1pt, draw=gray!10},
            major grid style={line width=.2pt,draw=gray!50},
            every axis/.append style={
                x label style={
                    font=\fontsize{8}{8}\selectfont,
                    at={(axis description cs:0.5,-0.1)},
                    },
                y label style={
                    font=\fontsize{8}{8}\selectfont,
                    at={(axis description cs:-0.1,0.5)},
                    },
                x tick label style={
                    font=\fontsize{8}{8}\selectfont,
                    /pgf/number format/.cd,
                    fixed,
                    fixed zerofill,
                    precision=2,
                    /tikz/.cd
                    },
                y tick label style={
                    font=\fontsize{8}{8}\selectfont,
                    /pgf/number format/.cd,
                    fixed,
                    fixed zerofill,
                    precision=1,
                    /tikz/.cd
                    },
            },
        }
        \begin{axis}[
            ymode=log,
            log ticks with fixed point,
        ]
        \addplot[no markers, gray] table [x=Step, y=DDPG, col sep=comma] {Data/conv_ch_code.csv};
        \addlegendentry{DDPG}
        
        \addplot[no markers, cyan] table [x=Step, y=REINFORCE, col sep=comma] {Data/conv_ch_code.csv};
        \addlegendentry{REINFORCE}
        
        \addplot[no markers, magenta] table [x=Step, y=A2C, col sep=comma] {Data/conv_ch_code.csv};
        \addlegendentry{Actor-Critic}
        \end{axis}
    \end{tikzpicture}
  \caption{Convergence behavior for the joint channel coding and modulation problem in an AWGN channel.}
  \label{fig:ch_coding_convergence}
\end{minipage}
\vspace{-0.8cm}
\end{figure}

In the BN channel case (Fig. \ref{fig:bn_grid_world}), similar observations can be made compared to the AWGN case. 
The biggest difference is that we see a larger performance improvement over the separation case when using our proposed framework than in the AWGN case.
This is particularly obvious when using BPSK modulation, where the gap between the BPSK results for the joint learning and communication scheme and those from the separate learning and communication is larger compared to the AWGN channel case.
This shows that in this more challenging channel scenario, the proposed framework is better able to adjust jointly the policy and the communication scheme to meet the conditions of the channel.
It also again highlights the fact that the Shannon capacity is not the most important metric for this problem as the expected SNR is not significantly less due to the burst noise but we observe an even more pronounced improvement using the proposed schemes over the separation schemes.

In Figs. \ref{fig:bsc_grid_world}, \ref{fig:awgn_grid_world} and \ref{fig:bn_grid_world}, it can be seen that when the grid world itself is noisy (i.e., $\delta>0$), the agents are still able to collaborate, albeit at the cost of higher average steps required to reach the treasure. 
The convergence of the number of steps used to reach the treasure for each channel scenario is shown in Fig. \ref{fig:mdp_convergence}. 
The slow convergence for the BSC channel indicates the difficulty of learning a binary code for this channel.
We also study the effect of the bandwidth $M$ on the performance. 
In Fig. \ref{fig:bw_affec}, we present the average number of steps required for channel bandwidths $M=7$ and $M=10$. 
As expected, increasing the channel bandwidth reduces the average number of steps for the scout to reach the treasure. 
The gain is particularly significant for BPSK at the low SNR regime as the guide is better able to protect the information conveyed against the channel noise thanks to the increased bandwidth.

Next, we present the results for the joint channel coding and modulation problem. 
Fig. \ref{fig:ch_coding_bler} shows the BLER performance obtained by BPSK modulation and Hamming (7,4) code, our DDPG transmitter described in Section \ref{subsec:eg_prob_channel_coding}, the one proposed by \cite{aoudia_model-free_2019}, and the proposed approach using an additional critic, labeled as ``Hamming (7,4)", ``DDPG", ``REINFORCE", and ``Actor-Critic", respectively.
It can be seen that the learning approaches (DDPG, REINFORCE and Actor-Critic) perform better than the Hamming (7,4) code. 
Additionally, stochastic policy algorithms (REINFORCE and Actor-Critic) perform better than DDPG.
This is likely due to the limitations of DDPG, as in practice, criterion 1) of Theorem \ref{thm:ddpg_compatibility} is often not satisfied.
Lastly, we show that we can improve upon the algorithm proposed in \cite{aoudia_model-free_2019} by adding an additional critic that reduces the variance of the policy gradients; and therefore, learns a better policy.
The results obtained by the actor-critic algorithm are superior to those from the REINFORCE algorithm, especially in the higher SNR regime.
On average, the learning-based results are better than the Hamming (7,4) performance by $1.24$, $2.58$ and $3.70$ dB for DDPG, REINFORCE and Actor-Critic, respectively.

\begin{figure} 
    \centering
  \subfloat[$p_b=0.1$ \label{subfig:ch_coding_bn01}]{%
    \begin{tikzpicture}
        \pgfplotsset{
            legend style={
                font=\fontsize{6}{6}\selectfont,
                at={(0.4,.35)},
                anchor=north east,
            },
            height=0.4\linewidth,
            width=0.5\linewidth,
            xmin=0,
            xmax=5,
            ymin=0.01,
            ymax=0.4,
            xlabel={SNR (dB)},
            ylabel={BLER},
            grid=both,
            grid style={line width=.1pt, draw=gray!10},
            major grid style={line width=.2pt,draw=gray!50},
            every axis/.append style={
                x label style={
                    font=\fontsize{8}{8}\selectfont,
                    at={(axis description cs:0.5,-0.1)},
                    },
                y label style={
                    font=\fontsize{8}{8}\selectfont,
                    at={(axis description cs:-0.1,0.5)},
                    },
                x tick label style={
                    font=\fontsize{8}{8}\selectfont,
                    /pgf/number format/.cd,
                    fixed,
                    fixed zerofill,
                    precision=2,
                    /tikz/.cd
                    },
                y tick label style={
                    font=\fontsize{8}{8}\selectfont,
                    /pgf/number format/.cd,
                    fixed,
                    fixed zerofill,
                    precision=1,
                    /tikz/.cd
                    },
            }
        }
        \begin{axis}[
            ymode=log,
            log ticks with fixed point,
        ]
        
        \addplot[blue, solid, line width=0.9pt, mark=triangle*, mark options={fill=blue, scale=1.5, solid}] 
                table [x=SNR, y=HAMMING, col sep=comma] {Data/ch_coding_bn01.csv};
        \addlegendentry{HAMMING}
        
        \addplot[gray, dashed, line width=1.2pt, mark=square*, mark options={fill=gray, solid, scale=1.}] 
                table [x=SNR, y=DDPG, col sep=comma] {Data/ch_coding_bn01.csv};
        \addlegendentry{DDPG}
        
        \addplot[cyan, solid, line width=0.9pt, mark=*, mark options={fill=cyan, scale=1.}] 
                table [x=SNR, y=REINFORCE, col sep=comma] {Data/ch_coding_bn01.csv};
        \addlegendentry{REINFORCE}
        
        \addplot[magenta, dashed, line width=1.2pt, mark=x, mark options={fill=magenta, scale=1.5, solid}] 
                table [x=SNR, y=A2C, col sep=comma] {Data/ch_coding_bn01.csv};
        \addlegendentry{Actor-Critic}
        
        \end{axis}
        \end{tikzpicture}
       }
  \subfloat[$p_b=0.2$ \label{subfig:ch_coding_bn02}]{%
    \begin{tikzpicture}
        \pgfplotsset{
            legend style={
                font=\fontsize{6}{6}\selectfont,
                at={(0.4,.35)},
                anchor=north east,
            },
            height=0.4\linewidth,
            width=0.5\linewidth,
            xmin=0,
            xmax=5,
            ymin=0.05,
            ymax=0.5,
            xlabel={SNR (dB)},
            ylabel={BLER},
            grid=both,
            grid style={line width=.1pt, draw=gray!10},
            major grid style={line width=.2pt,draw=gray!50},
            every axis/.append style={
                x label style={
                    font=\fontsize{8}{8}\selectfont,
                    at={(axis description cs:0.5,-0.1)},
                    },
                y label style={
                    font=\fontsize{8}{8}\selectfont,
                    at={(axis description cs:-0.15,0.5)},
                    },
                x tick label style={
                    font=\fontsize{8}{8}\selectfont,
                    /pgf/number format/.cd,
                    fixed,
                    fixed zerofill,
                    precision=2,
                    /tikz/.cd
                    },
                y tick label style={
                    font=\fontsize{8}{8}\selectfont,
                    /pgf/number format/.cd,
                    fixed,
                    fixed zerofill,
                    precision=2,
                    /tikz/.cd
                    },
            }
        }
        \begin{axis}[
            ymode=log,
            log ticks with fixed point,
        ]
        
        \addplot[blue, solid, line width=0.9pt, mark=triangle*, mark options={fill=blue, scale=1.5, solid}] 
                table [x=SNR, y=HAMMING, col sep=comma] {Data/ch_coding_bn02.csv};
        \addlegendentry{HAMMING}
        
        \addplot[gray, dashed, line width=1.2pt, mark=square*, mark options={fill=gray, solid, scale=1.}] 
                table [x=SNR, y=DDPG, col sep=comma] {Data/ch_coding_bn02.csv};
        \addlegendentry{DDPG}
        
        \addplot[cyan, solid, line width=0.9pt, mark=*, mark options={fill=cyan, scale=1.}] 
                table [x=SNR, y=REINFORCE, col sep=comma] {Data/ch_coding_bn02.csv};
        \addlegendentry{REINFORCE}
        
        \addplot[magenta, dashed, line width=1.2pt, mark=x, mark options={fill=orange, scale=1.5, solid}] 
                table [x=SNR, y=A2C, col sep=comma] {Data/ch_coding_bn02.csv};
        \addlegendentry{Actor-Critic}
        
        \end{axis}
        \end{tikzpicture}
        }
  \caption{Comparison of the agents jointly trained to collaborate and communicate over an BN channel to separate learning and communications with a (7,4) Hamming code.}
  \label{fig:ch_coding_bler_bn} 
\vspace{-1.0cm}
\end{figure}
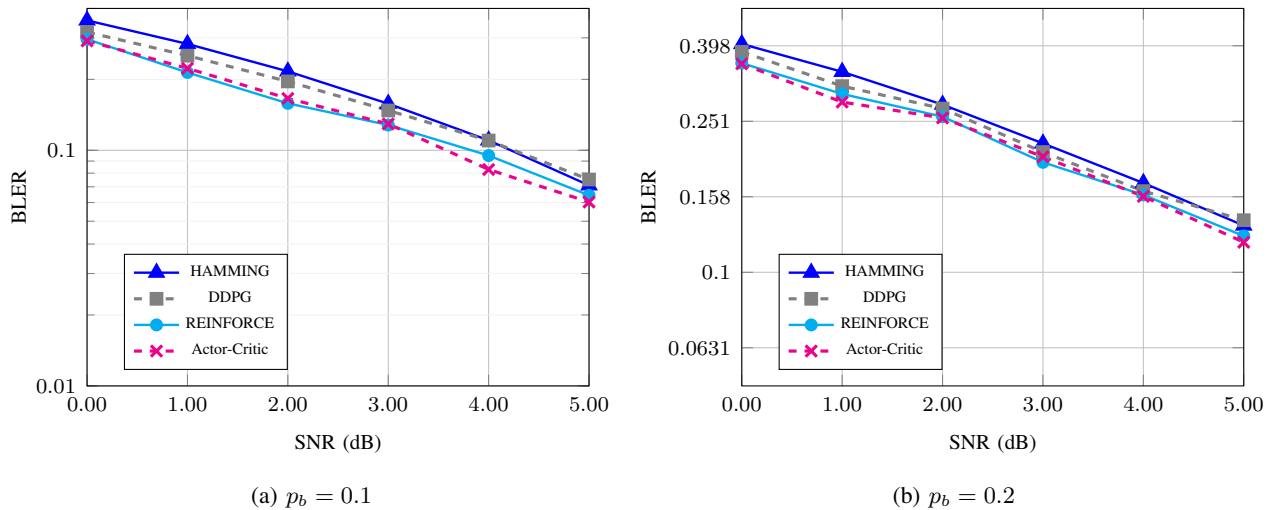

When considering the BN channel case, as shown in Fig. \ref{fig:ch_coding_bler_bn}, while the BLER increases due to the increased noise for all the schemes, we still see improved performance with the learning algorithms. 
Fig. \ref{fig:ch_coding_convergence} shows the convergence behavior of different learning algorithms for 5dB channel SNR.
We can see that the actor-critic algorithm converges the quickest and achieves the lowest BLER, while REINFORCE converges the slowest but achieves lower BLER than DDPG at the end of training.
This is in accordance with the BLER performance observed in Fig. \ref{fig:ch_coding_bler}.
We reiterate that the joint channel coding and modulation problem studied from the perspective of supervised learning in \cite{aoudia_model-free_2019} is indeed a special case of the joint learning and communication framework we presented in Section \ref{sec:problem_formulation} from a MARL perspective, and can be solved using a myriad of algorithms from the RL literature.

Lastly, we note that due to the simplicity of our network architecture, the computation complexity of our models is not significantly more than the separation based results we present herein.
The average computation time for encoding and decoding using our proposed DRL solution is approximately $323\mu s$ compared to $286 \mu s$ for the separate learning and communication case with a Hamming (7,4) code, using an Intel Core i9 processor.
This corresponds to roughly 13\% increase in computation time, which is modest considering the performance gains observed in both the guided robot problem and the joint channel coding and modulation problem.

\begin{Remark}
We note that both the grid world problem and the channel coding and modulation problems are POMDPs. Therefore, recurrent neural networks (RNNs), such as long-short term memory (LSTM) \cite{hochreiter_lstm_1997} networks, should provide performance improvements as the cell states can act as belief propagation. However, in our initial simulations, we were not able to observe such improvements, although this is likely to be due to the limitations of our architectures.
\end{Remark}

\begin{Remark}
Even though we have only considered the channel modulation and coding problem in this paper due to lack of space, our framework can also be reduced to the source coding and joint source-channel coding problems by changing the reward function. If we consider an error-free channel with binary inputs and outputs, and let the reward depend on the average distortion between the $B$-length source sequence observed by agent 1 and its reconstruction generated by agent 2 as its action, we recover the lossy source coding problem, where the length-$B$ sequence is compressed into $M$ bits. If we instead consider a noisy channel in between the two agents, we recover the joint source-channel coding problem with an unknown channel model. 
\end{Remark}



\section{Conclusion}
\label{sec:conclusions}

In this paper, we have proposed a comprehensive framework that jointly considers the learning and communication problems in collaborative MARL over noisy channels. 
Specifically, we consider a MA-POMDP where agents can exchange messages with each other over a noisy channel in order to improve the shared total long-term average reward.
By considering the noisy channel as part of the environment dynamics and the message each agent sends as part of its action, the agents not only learn to collaborate with each other via communications but also learn to communicate ``effectively".  
This corresponds to ``level C'' of Shannon and Weaver's organization of the communication problems in \cite{ShannonWeaver49}, which seeks to answer the question ``How effectively does the received meaning affect conduct in the desired way?".
We show that by jointly considering learning and communications in this framework, the learned joint policy of all the agents is superior to that obtained by treating the communication and the underlying MARL problem separately. 
We emphasize that the latter is the conventional approach when the MARL solutions obtained in the machine learning literature assume error-free communication links are employed in practice when autonomous vehicles or robots communicate over noisy wireless links to achieve the desired coordination and cooperation. 
We demonstrate via numerical examples that the policies learned from our joint approach produce higher average rewards than those where separate learning and communication is employed.
We also show that the proposed framework is a generalization of most of the communication problems that have been traditionally studied in the literature, corresponding to ``level A'' as described by Shannon and Weaver. 
This formulation opens the door to employing available numerical MARL techniques, such as the actor-critic framework, for the design of channel modulation and coding schemes for communication over unknown channels. 
We believe this is a very powerful framework, which has many real world applications, and can greatly benefit from the fast developing algorithms in the MARL literature to design novel communication codes and protocols, particularly with the goal of enabling collaboration and cooperation among distributed agents.

\bibliographystyle{ieeetr}
\bibliography{bare_jrnl.bib}

\end{document}